\documentclass[10pt,twocolumn,aps,prb,balancelastpage]{revtex4-1}
\usepackage[latin9]{inputenc}
\usepackage{color}
\usepackage{verbatim}
\usepackage{amsmath}
\usepackage{amssymb}
\usepackage{graphicx}
\usepackage{esint}
\usepackage[unicode=true,pdfusetitle,
 bookmarks=true,bookmarksnumbered=false,bookmarksopen=false,
 breaklinks=false,pdfborder={0 0 1},backref=false,colorlinks=true]{hyperref}
\usepackage{breakurl}
\usepackage{epstopdf}

\makeatletter

\@ifundefined{textcolor}{}
{%
 \definecolor{BLACK}{gray}{0}
 \definecolor{WHITE}{gray}{1}
 \definecolor{RED}{rgb}{1,0,0}
 \definecolor{GREEN}{rgb}{0,1,0}
 \definecolor{BLUE}{rgb}{0,0,1}
 \definecolor{CYAN}{cmyk}{1,0,0,0}
 \definecolor{MAGENTA}{cmyk}{0,1,0,0}
 \definecolor{YELLOW}{cmyk}{0,0,1,0}
}

\usepackage{subfigure}\usepackage{bm}

\newcommand{\bv}[1]{\mathbf{#1}}

\makeatother

\begin{document}

\title{Quantum Oscillations in Weyl and Dirac Semimetal Ultra-Thin Films}

\author{Daniel Bulmash, Xiao-Liang Qi}

\affiliation{Department of Physics, Stanford University, Stanford, California 94305-4045, USA}

\date{\today}
\begin{abstract}
We show that a thin film of Weyl or Dirac semimetal with a strong in-plane 
magnetic field becomes a novel two-dimensional Fermi liquid with interesting 
properties. The Fermi surface in this system is strongly anisotropic, 
which originates from a combination of chiral bulk channels and the Fermi arcs. The 
area enclosed by the Fermi surface depends strongly on the in-plane magnetic field 
component parallel to the Weyl/Dirac node splitting, which leads to unusual 
behavior in quantum oscillations when the magnetic field is tilted out of the 
plane. We estimate the oscillation frequencies and the regimes where such 
effects could be seen in Cd$_3$As$_2$, Na$_3$Bi, and TaAs. 
\end{abstract}
\maketitle

 Weyl \cite{Murakami2007,PyrochloreWeyl,WeylMultiLayer} and Dirac \cite{YoungDiracSemimetal} semimetals (WSMs and DSMs respectively) are new three dimensional phases which have recently generated a great deal of interest as the first examples of topological phases of gapless systems. A WSM has topologically robust linear band touchings at discrete points, called Weyl nodes, in the bulk Brillouin zone and surface ``Fermi arcs" which connect the projections of the Weyl nodes to the surface Brillouin zone. WSMs are predicted to have many novel transport properties related to the chiral anomaly \cite{ZyuninBurkovWeylTheta,SonSpivakWeylAnomaly,QiWeylAnomaly,HosurWeylReview}. DSMs are WSMs where several Weyl nodes overlap in momentum space and are protected by symmetries.

Since the prediction and discovery of the DSMs Na$_3$Bi \cite{WangA3Bi,Xu2013,Liu2014} and Cd$_3$As$_2$ \cite{WangCd3As2,LiuCd3As2,Neupane2014,BorisenkoCd3As2} along with the TaAs class of WSMs \cite{TaAsPredictionPRX,TaAsPrediction,PrincetonTaAs,CASTaAs,YangTaAs}, a wide range of experiments, particularly in transport, have found unusual behavior in these materials. Negative longitudinal magnetoresistance, suggestive of the chiral anomaly, appears even far from the quantum limit, and linear transverse magnetoresistance is widespread among these materials \cite{BiSbKimMagnetoTransportExpt,LiCd3As2NegativeMR,LiangCd3As2,HeCd3As2, XiongCurrentPlume,ZhangTaAsTransport}. Recently, experiments have begun to directly probe the quantum and thin film limits\cite{ZhangTaAsTransport,MollNbAs,LiuGatedCd3As2}.

\begin{figure}
\subfigure[ ]{
\includegraphics[width=3.5cm]{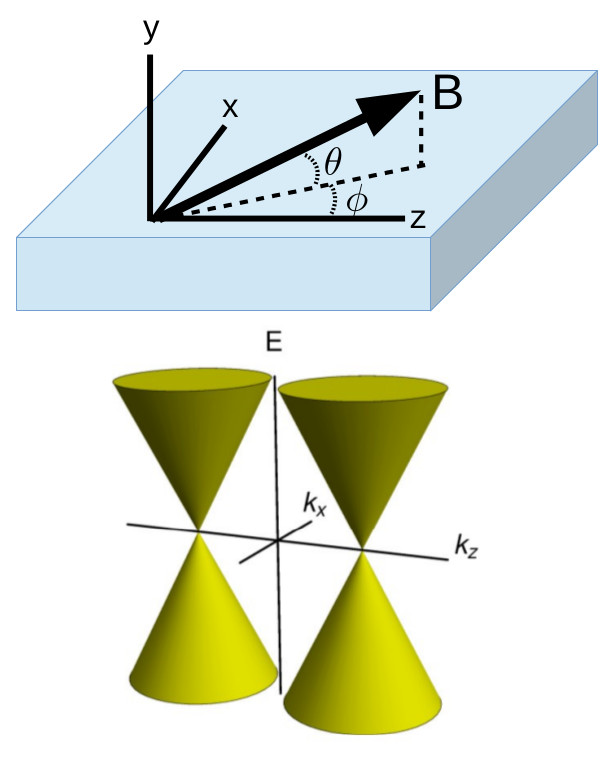}
\label{fig:geometry}
}
\subfigure[ ]{
\includegraphics[width=3.5cm]{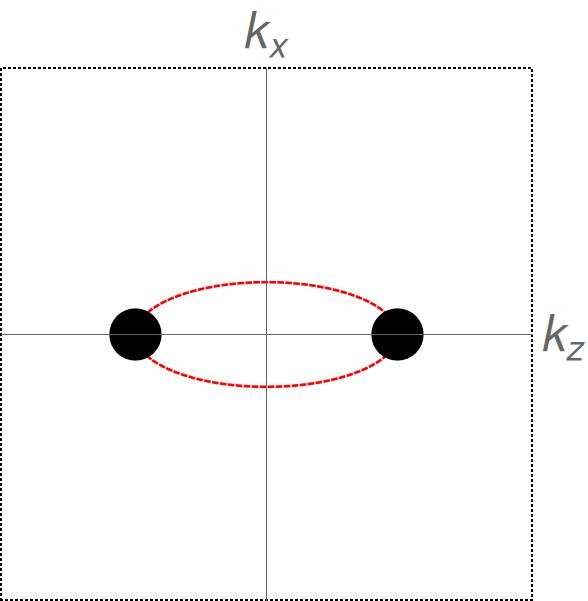}
\label{fig:BulkFS}
}
\subfigure[ ]{
\includegraphics[width=4cm]{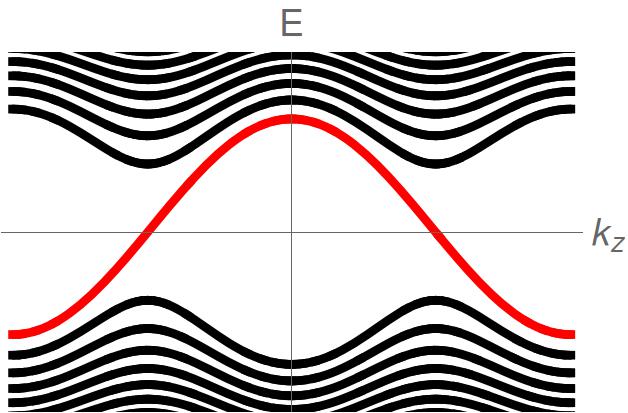}
\label{fig:bulkDispersion}
}
\subfigure[ ]{
\includegraphics[width=3.5cm]{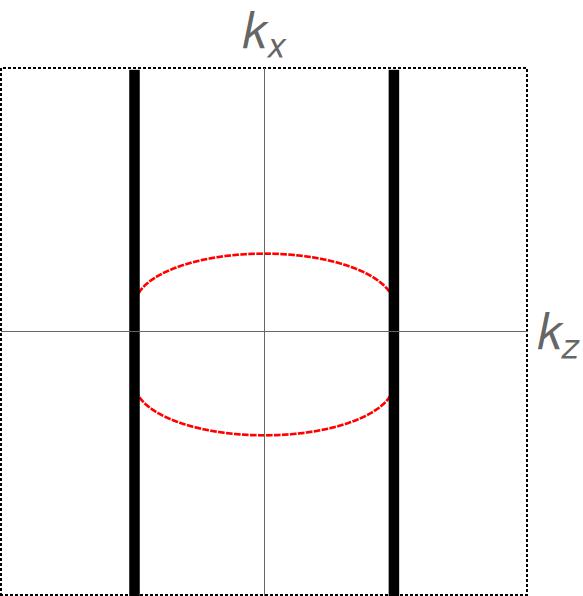}
\label{fig:BulkFS_WithB}
}
\caption{(a) Basic geometry considered in this paper. The sample is a thin film in $y$ with magnetic field primarily along the $z$ direction (upper figure) and the Weyl point splitting along $k_z$ (lower figure). (b) Schematic Fermi surface of a 2-node WSM in a thick slab geometry. The bulk contribution is in black and the surface Fermi arcs are dashed red. (c) Schematic bulk dispersion of a WSM in a strong magnetic field. The zeroth Landau level is shown in red and disperses chirally near each Weyl point. (d) Schematic Fermi surface of a thick 2-node WSM in a strong in-plane field. The color scheme is the same as (b). The precise locations of the Fermi arcs depend on thickness and field strength.}
\end{figure}

In this paper, we show that the unique properties of WSMs and DSMs have new consequences in the thin film limit. With an in-plane magnetic field along suitable directions, we show that a thin film of WSM or DSM becomes a two-dimensional Fermi liquid with a highly anisotropic and magnetic-field tunable Fermi surface. Our setup is shown schematically in Fig. \ref{fig:geometry}. This two-dimensional Fermi surface emerges from a combination of the surface Fermi arcs and the chiral channels in a bulk WSM or DSM with a magnetic field. Our main result is that the shape of this Fermi surface is tuned not only by the shape of the Fermi arcs but also the in-plane mangetic field. This tunability, which is not present for a solely out-of-plane field, can be probed directly by quantum oscillations in a magnetic field with an out-of-plane component. The most drastic contrast to ordinary  two-dimensional metals occurs when the Fermi arcs have no curvature. In this case, the density of states (DOS) oscillates as a function of field angle at fixed field strength, but not as a function of field strength at fixed angle. As concrete predictions for future experiments, we estimate the parameters of these novel quantum oscillations in Cd$_3$As$_2$, Na$_3$Bi, and TaAs. The unusual origin of the Fermi surface in this system may have other consequences due to the strong anisotropy of electron wavefunctions on the Fermi surface. We will discuss these possibilities at the end of the paper.

Recently, quantum oscillations coming from the area enclosed by the Fermi arcs were predicted in thin films of WSM with perpendicular magnetic field\cite{PotterFermiArcOsc,ZhangWSMQOs}; evidence for this prediction was recently observed experimentally \cite{MollCd3As2Osc}. Our results cross over to but are in a different regime from those of Ref. \onlinecite{PotterFermiArcOsc} and its interacting weak-field generalization \cite{GorbarWSMQOs} due to the strong in-plane field. Accordingly, the features of the quantum oscillations in these two different setups are also qualitatively different. For quantum oscillations as a function of the out-of-plane field component, Refs. \onlinecite{PotterFermiArcOsc,ZhangWSMQOs,GorbarWSMQOs} predict no dependence on the in-plane component in the non-interacting case while in our results, that component tunes the oscillation frequency.

\textit{Emergent Fermi surface:} For simplicity, we consider a minimal model of a WSM with two Weyl nodes at the wavevectors $\bv{k} = \pm k_W\bv{\hat{z}}$ in the slab geometry shown in Fig. \ref{fig:geometry}. Before studying the thin-film limit, we first consider the properties of a thick film. At zero field and finite but small chemical potential, the Fermi surface of a thick film is as shown in Fig. \ref{fig:BulkFS}; it consists of two small, spherical bulk Fermi surfaces connected by Fermi arcs on opposite real space surfaces. Adding a magnetic field $B$ in the $\bv{\hat{z}}$ direction, we can choose a Landau gauge $\bv{A} = -eBy\bv{\hat{x}}$ for the vector potential such that full in-plane translation symmetry is preserved after Peierls substitution. The magnetic field causes the formation of Landau levels, which quenches the momentum $k_x$ and locks the eigenfunctions' average $y$ position to $k_x$ via $\langle y \rangle = k_x l_B^2$. Here $l_B = \sqrt{\hbar/eB}$ is the magnetic length. However, the Landau levels still disperse in $k_z$. In particular, near a Weyl point, where we will take for simplicity the effective Hamiltonian to be $H = \hbar v_1(k_x \sigma_x + k_y\sigma_y) + \hbar v_2 k_z\sigma_z$ (here $\sigma_i$ are the Pauli matrices and $\bv{k}$ is measured from the Weyl point), it is easy to show that there is a single zeroth Landau level (ZLL) with dispersion
\begin{equation}
E_0(k_x,k_z) = -\hbar v_2k_z
\end{equation}

Since the Weyl points come in pairs of opposite chirality, the sign of $v_2$ must be different at the two Weyl points, leading to a dispersion like that in Fig. \ref{fig:bulkDispersion}. If the chemical potential is small enough that it only crosses the ZLL, then such a dispersion leads to a quasi-1D Fermi surface shown in Fig. \ref{fig:BulkFS_WithB} of width roughly equal to $2k_W$. Note that the Fermi arcs still exist, but we will see that they play an unimportant role in the thick limit.

Our key observation, however, is in the thin film limit. To investigate this limit, we diagonalized a minimal 2-band lattice model
\begin{align}
H = &2t\sin k_x \sigma_x + 2t \sin k_y \sigma_y \nonumber\\
 &+\left(M+2C \cos k_z + 2A(2-\cos k_x - \cos k_y)\right)\sigma_z
 \label{eqn:2BandH}
\end{align}
and included the magnetic field via Peierls substitution. (We have set the lattice constant $a=1$.) The bulk model has Weyl nodes at $k_x=k_y=0$, $k_z = \pm \cos^{-1}(-M/2C)$, and these are the only Weyl nodes if $|M|+|2C| < 4|A|$ (which we will always assume).

\begin{figure}
\subfigure[ ]{
\includegraphics[width=4cm]{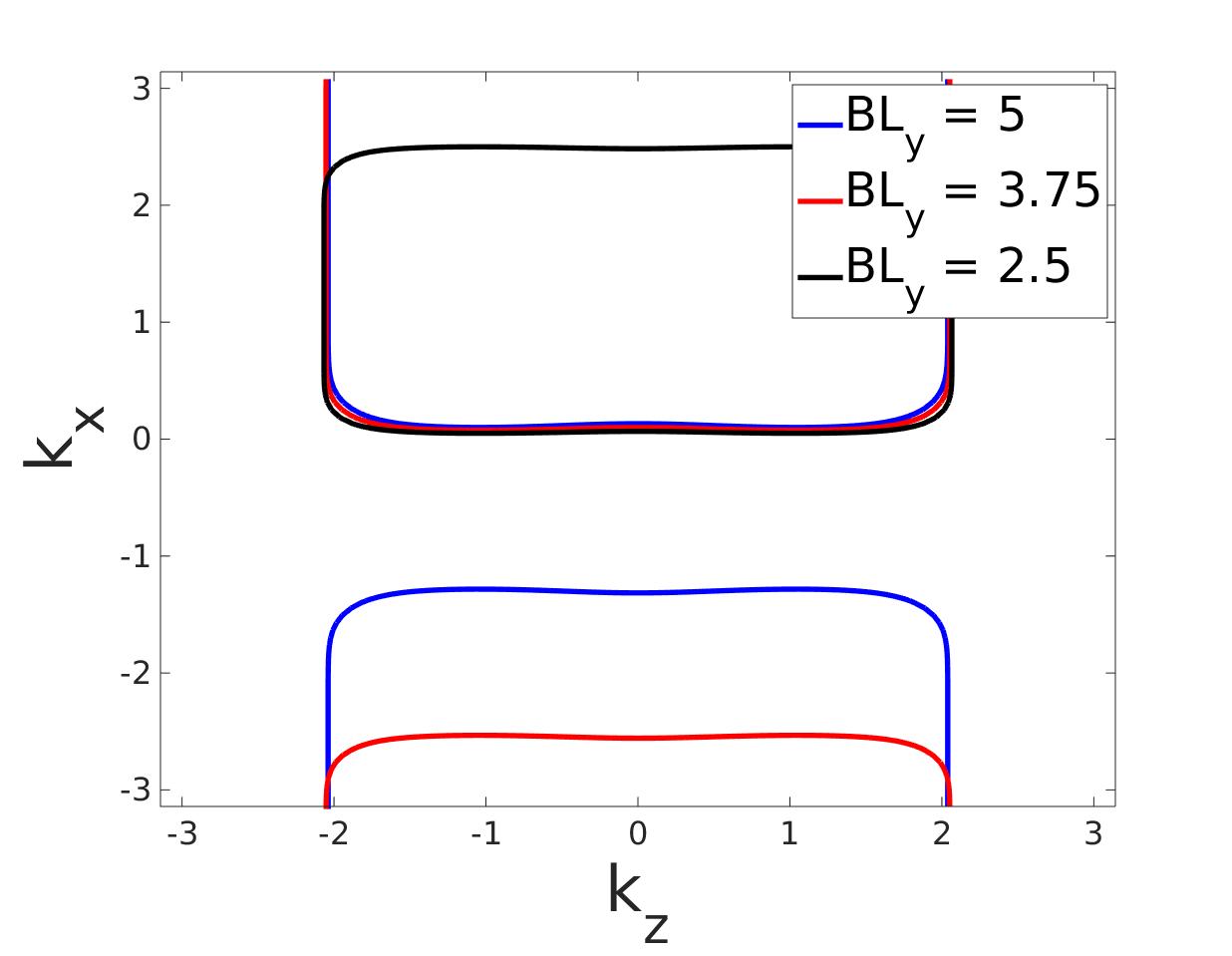}
\label{fig:FSEvolution}}
\subfigure[ ]{
\includegraphics[width=4cm]{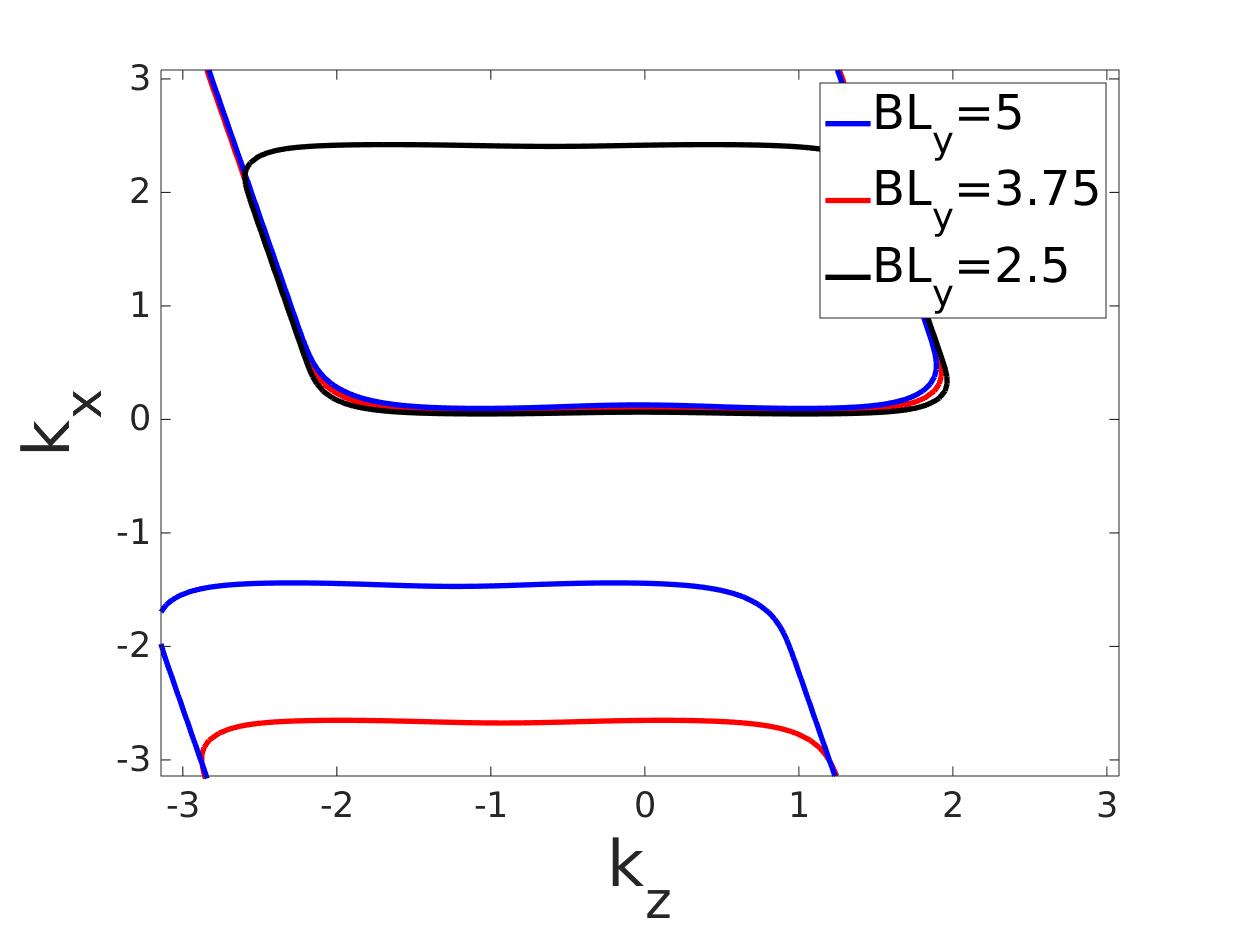}
\label{fig:skewedFS}
}
\caption{Numerically calculated evolution of the 2D Fermi surface of Eq. (\ref{eqn:2BandH}). We set $e=1$ and the lattice constant $a=1$. Parameters: $t=A=1$, $M=C=-1$. (a) Evolution in $B_zL_y$ for a purely $z$-direction field. The length of the Fermi surface in $k_x$ is proportional to $BL_y$, and the width in $k_z$ is set by the Weyl point separation $\cos^{-1}(-M/2C)$. We checked (not shown) that fixing $BL_y$ and changing $L_y$ only changes the curvature of the Fermi arcs. (b) Fermi surface with an angled field with $\theta = 0.08\pi$ from the $z$-axis and $BL_y = 2.5$. Comparing with the analogous curve in (a) we see that the vertical portions of the Fermi surface skew by the angle $\theta$.}
\end{figure}

In this case, at zero field, the Fermi surface is similar to the thick case in Fig. \ref{fig:BulkFS}. However, the picture in the quantum limit of a $z$-direction magnetic field is quite different from Fig. \ref{fig:BulkFS_WithB}. As $k_x$ increases, position/momentum locking causes the average $y$ position to increase as well. Therefore, when $k_x \approx 0$ or $L_y/l_B^2$ with $L_y$ the sample thickness, the eigenstates reach a surface and thus must disperse along $k_x$. But we already know that there are other gapless modes at the surface, namely the Fermi arcs. Since the Fermi arcs can be thought of as quantum anomalous Hall edge states (at fixed $k_z$), we expect that the bulk Fermi surface merges with the Fermi arcs, leading to a closed, two-dimensional Fermi surface shown in Fig. \ref{fig:FSEvolution}. The existence of this closed 2D Fermi surface in the quantum limit of a WSM is our primary result.

In fact the same effect can occur in standard metals; rather than Fermi arcs, the surface modes can come from band bending effects, for example. However, because our picture only makes sense in the quantum limit, the carrier density must be very low, so the width of this Fermi surface in a metal will be very small. By contrast, in a WSM the $B=0$ carrier density can even be zero while still maintaining a finite width $2k_W$ of the 2D Fermi surface. For example, in a quadratic band with isotropic effective mass, if we want the quantum limit to occur at an energy where the 2D Fermi surface has $k_F = 0.1$ \AA$^{-1}$ (which is the order of magnitude of the Weyl point splittings in TaAs \cite{CASTaAs,PrincetonTaAs}), then an unphysically large field of $360$ T is required. To match the $\sim 0.03$ \AA$^{-1}$ Weyl point splitting in Cd$_3$As$_2$ \cite{LiangCd3As2} a more reasonable but still large field of about $30$ T would be required.

We must point out that due to the quenching of $k_x$, there is perfect nesting at, for example, $\bv{q} = 2k_W\bv{\hat{z}}$. As such, one might expect a charge density wave (CDW) instability of the 2D Fermi surface, as has been predicted theoretically \cite{Gusynin1996}. To our knowledge this effect has not been seen in any system in the quantum limit. A possible reason is that, due to the estimations of the previous paragraph, the dominant instability (assuming $q_x = 0$) would be at a very small wavevector in a metal, unlike in a WSM. The CDW instability in WSM thin films is by itself an interesting topic, but for the rest of this paper we will make the assumption that there is no CDW and that the Fermi surface is robust.

The existence of this Fermi surface implies that if we add a magnetic field in the $y$ direction $B_y \ll B_z$, then there will be quantum oscillations as a function of $B_y$. However, an unusual feature of this Fermi surface is that its length in $k_x$ is controlled by $L_y/l_B^2 \propto L_y B_z$, as demonstrated by the different curves in Fig. \ref{fig:FSEvolution}. Therefore, since the area of this Fermi surface is tuned by the in-plane magnetic field, the frequency of quantum oscillations in $B_y$ will also depend on $B_z$. We will shortly discuss this point in some depth.

However, before looking at quantum oscillations (i.e. an out-of-plane field), we should understand how this Fermi surface evolves when the field is not perfectly aligned with the Weyl point separation. First consider an in-plane rotation $\phi$ of the magnetic field. Then dispersion occurs along the direction of the field, so the vertical portions of the Fermi surface in Fig. \ref{fig:FSEvolution} should simply skew to be perpendicular to the field. Their lengths $L_y/l_B^2$ should also be preserved, as $y$ position is locked to the component of momentum perpendicular to the field. The net result is a reduction in Fermi surface area by $|\cos \phi|$, which we see in the numerics in Fig. \ref{fig:skewedFS}.

We also need to understand what happens if the Weyl point separation has a significant component in the $y$ direction. For this we added a term $-2t\alpha \sin k_z \sigma_y + (-2A+A\sqrt{4+\alpha^2(M^2-4)})\sigma_z$ to the Hamiltonian, where $\alpha \in [0,4/(4-M^2)]$. The first term shifts the Weyl points to a nonzero $k_y$, and the second term is used to keep the $k_z$ separation of the Weyl points fixed. Numerically, we find that the result is to amplify finite size effects in the $k_z$ width of the Fermi surface. This effect is small, however; for $\alpha = 0.9$ when $M=-t$, the width only changes by about $10$\% between the bulk limit and $L_y = 50$. We will thus neglect these effects from now on.

\textit{Quantum oscillations:} To predict observable properties of the field-tuned Fermi surface, we study quantum oscillations by applying in addition a perpendicular magnetic field. Suppose we add a small $B_y \ll B_z$. Then the Bohr-Sommerfeld quantization rule says that Landau levels are at energies where
\begin{equation}
\frac{1}{B_y} = (n+\lambda)\frac{2\pi e}{\hbar A_{FS}}
\label{eqn:BohrSommerfeld}
\end{equation}
where $\lambda$ is a dispersion-dependent constant, $n$ is an integer, and $A_{FS}$ is the Fermi surface area. The Fermi surface area can be estimated as the sum of two parts. One is a constant $\delta A$ that the Fermi arcs enclose due to their curvature when $B=0$ and the chemical potential is at the Weyl points; this contributes as expected in previous work\cite{PotterFermiArcOsc,ZhangWSMQOs}. The new piece of the Fermi surface, discussed in the previous section, is rectangular, with length $L_y/l_{B_z}^2$ and width $2k_W$. Plugging into Eq. (\ref{eqn:BohrSommerfeld}),
\begin{equation}
\frac{1}{B_y} =\frac{2\pi e(n+\lambda)}{2k_W L_y e B_z + \hbar \delta A}
\label{eqn:ourOscillations}
\end{equation}

Eq. (\ref{eqn:ourOscillations}) is our main experimental prediction, valid for any Fermi arc configuration. It tells us that the frequency of quantum oscillations in $B_y$ is tuned by $B_z$. As particularly interesting special case, take the zero-curvature limit $\delta A \rightarrow 0$. Letting the field angle in the $yz$-plane be $\theta$ and the field angle in the $xz$-plane be $\phi$, as shown in Fig. \ref{fig:geometry}, Eq. (\ref{eqn:ourOscillations}) becomes
\begin{equation}
\cot \theta = \frac{\pi(n+\lambda)}{k_W L_y|\cos \phi|}
\label{eqn:angleOsc}
\end{equation}
Such quantum oscillations are qualitatively different from those in ordinary 2D or 3D systems because the oscillations occur as a function of field direction $\theta$, not total field strength, if $\phi$ is kept constant.

\begin{figure}
\includegraphics[width=9cm]{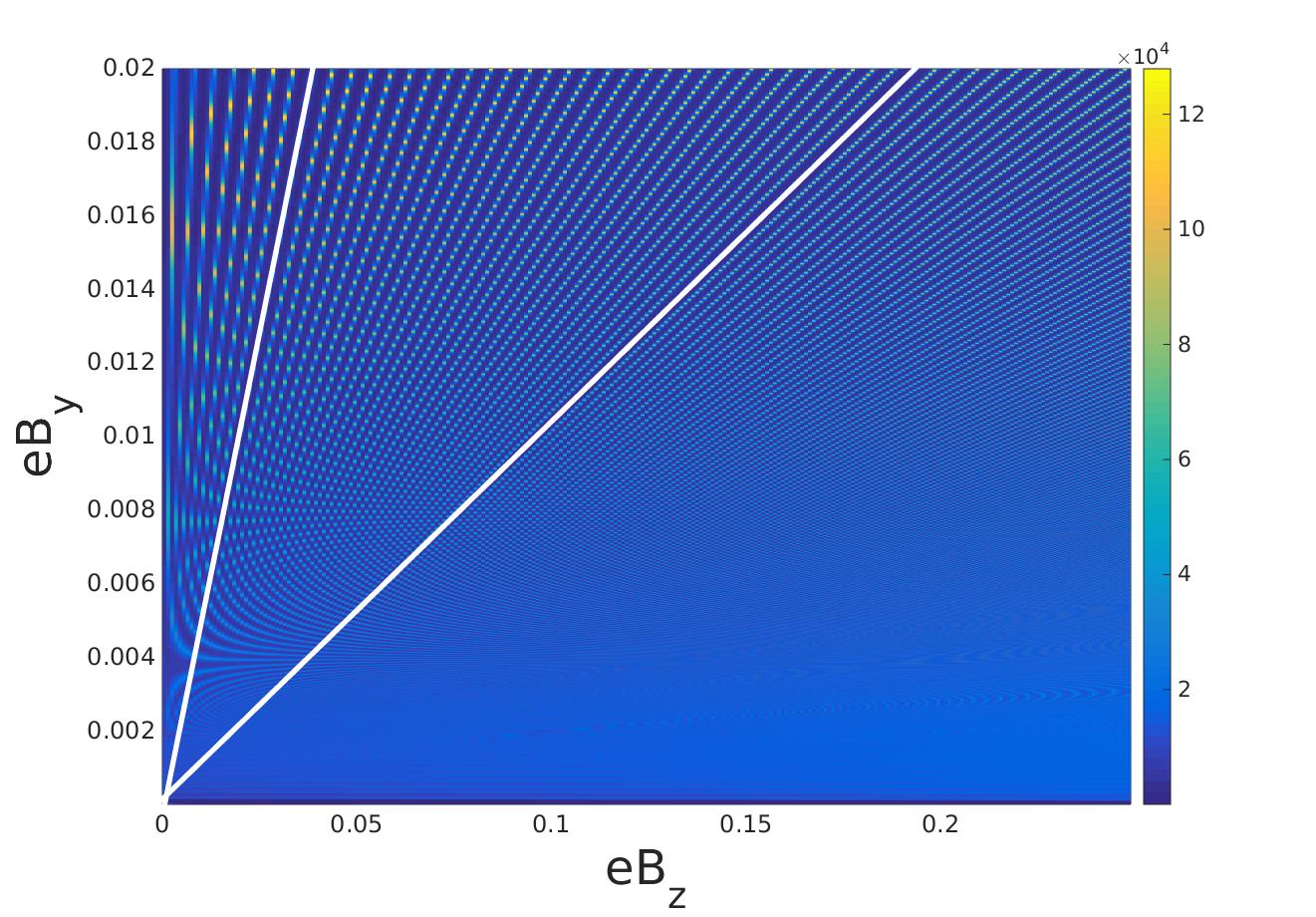}
\caption{Numerical density of states for the Hamiltonian in Eq. \ref{eqn:2BandH} in field as a function of $eB_z$ (in-plane, along the Weyl node splitting) and $eB_y$ (out of plane). Parameters: $t=0.7$, $M=0$, $A=1$, $C=-1$, $L_y = 10$, Green's function broadening $\delta = 10^{-3}$. The white lines are guides to the eye for Landau levels $n=8$ and $n=38$; they have slightly different $y$-intercepts due to finite size effects. The arc-like features at very small $B_z$ are artifacts due to aliasing.}
\label{fig:DOSnumerics}
\end{figure}

Eq. (\ref{eqn:angleOsc}) requires $\theta$ be small to maintain $B_y \ll B_z$. It should be noted that the frequency of oscillation in $\cot\theta$ only depends on the intrinsic parameter $k_WL_y$ of the WSM thin film. As $\theta$ increases, if we consider a finite zero-field carrier density\footnote{Recent work\cite{ZhangWSMQOs} shows that the results in Ref. \onlinecite{PotterFermiArcOsc} are valid only when the Fermi arcs enclose zero area when the chemical potential is at the Weyl nodes, and gives a more general result. The results of the present paper in the $\theta \rightarrow \pi/2$ limit are consistent with the generalization.}, we expect a crossover to the behavior in Ref. \onlinecite{PotterFermiArcOsc}, where the oscillation frequency in $1/B_y$ is independent of the applied magnetic field. These analytic results are explicitly verified by numerically computing the DOS in a generic magnetic field using an iterative Green's function method, the result of which is shown in Fig. \ref{fig:DOSnumerics}.

One important, natural question is whether we can make an arbitrarily large Fermi surface by moving to thicker and thicker samples. If $L_y/l_B^2 > 2\pi/a$, with $a$ the lattice constant in the $x$ direction, then the Fermi surface should wrap around the Brillouin zone. Because states in the ZLL which differ in $k_x$ by $2\pi/a$ are spatially separated in the $y$ direction by $2\pi l_B^2/a$, hybridization between them should be exponentially suppressed; indeed, we numerically find near-degeneracy. Naively, then, we could just increase the size of the Fermi surface without bound. However, we will show that increased sample thickness suppresses the amplitude of quantum oscillations in the presence of scattering or finite temperature.

First, recall that if the time it takes to traverse the Fermi surface is longer than the scattering time $\tau$, then the electron cannot make a full orbit around the Fermi surface. This amounts to the condition
\begin{equation}
2\frac{L_y/l_{B_z}^2+2k_W}{\dot{k}} = \frac{2L_yB_z}{v_FB_y}+\frac{4\hbar k_W}{ev_FB_y} \ll \tau
\label{eqn:lifetime}
\end{equation}
which is one limit on $L_y$.

\begin{figure}
\includegraphics[width=6cm]{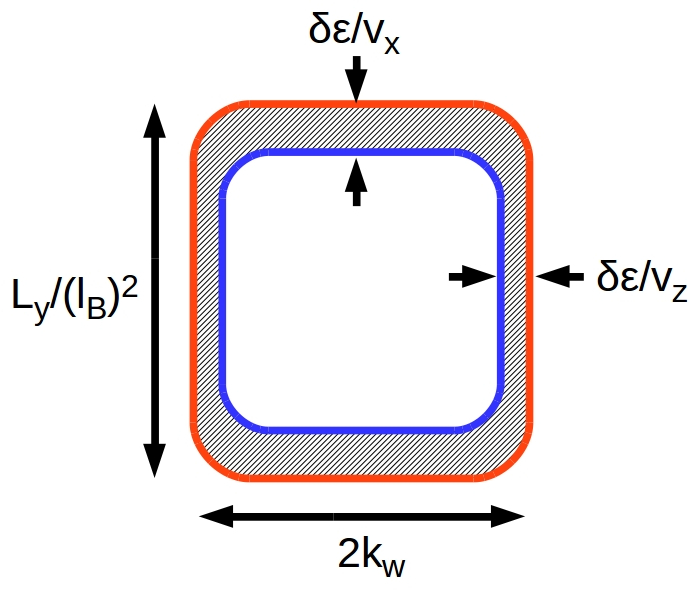}
\caption{Change in the 2D Fermi surface (blue) when the chemical potential is increased by $\delta \varepsilon$ (red). The change in Fermi surface area is the black hatched region, and can be estimated by taking the Fermi surface to be rectangular.}
\label{fig:LLSpacing}
\end{figure}

To understand how finite temperature limits $L_y$, we need to know what sets the Landau level gap for our 2D Fermi surface. We estimate this gap semiclassically. For the roughly rectangular Fermi surfaces in Fig. \ref{fig:FSEvolution}, the change in the area of the Fermi surface upon an increase of the chemical potential by $\delta \varepsilon$ can be estimated from Fig. \ref{fig:LLSpacing} to be
\begin{equation}
\delta A_{FS} \approx 2\left(\frac{\delta \varepsilon}{\hbar v_z}\right)\left(\frac{L_y}{l_{B_z}^2}\right)+2\left(\frac{\delta \varepsilon}{\hbar v_x}\right)(2k_W)
\label{eqn:deltaA}
\end{equation}
where $v_x$ is the Fermi arc Fermi velocity and $v_z$ is the bulk Fermi velocity in the $z$ direction. The quantization condition Eq. (\ref{eqn:BohrSommerfeld}) says that when $\delta \varepsilon$ is the Landau level spacing, $\delta A_{FS}$ obeys
\begin{equation}
\delta A_{FS} = \frac{2\pi e B_y}{\hbar}
\end{equation}
Substituting into Eq. (\ref{eqn:deltaA}) and solving, we find
\begin{equation}
\delta \varepsilon = \pi eB_y\left(\frac{eB_zL_y}{\hbar v_z}+\frac{2k_W}{v_x}\right)^{-1}
\label{eqn:LLSplitting}
\end{equation}
which is the temperature scale over which we can resolve the Landau level splitting. Note that we neglected Fermi surface curvature, which is a legitimate approximation as long as the film is not extremely thin. 

\textit{Estimations for real materials:} We now estimate the Fermi surface sizes for Cd$_3$As$_2$, Na$_3$Bi, and TaAs. For additional experimentally relevant estimations, such as where the thick limit occurs and detailed dependences of the frequencies on field angle, see the Appendix \ref{app:materials}.

Cd$_3$As$_2$ is a Dirac semimetal whose nodes are split along the $[001]$ plane, so each node contains both chiralities of Weyl points. As a result, the Dirac points are connected by two Fermi arcs per surface, and our picture yields two Fermi surfaces, a hole-like one and an electron-like one, whose difference in size is set by the chemical potential $\mu$. Taking them to be approximately rectangular as before and using parameters from Ref. \onlinecite{LiangCd3As2} (most importantly $2k_W = 0.03 \AA^{-1}$) we estimate
\begin{equation}
A_{FS} \sim \frac{e}{\hbar}\left(600 \text{ mT}\right)\left(1\pm\frac{\mu}{200 \text{ meV}} \right) \left(\frac{L_{y}}{1 \text{ nm}}\right) \left(\frac{B_{z}}{1 \text{ T}}\right)
\end{equation}
In fact, recent experiments\cite{LiuGatedCd3As2} were able to gate-tune a Cd$_3$As$_2$ thin film. They saw quantum oscillations at fixed field angle at some gate voltages. However, near what they identified as the Dirac point, they saw no contribution of the sort that we propose in Eq. (\ref{eqn:angleOsc}). This may be because their magnetic length was only 5 times smaller than the sample thickness, leading to considerable deformation of the emergent Fermi surface and the Landau level states. It may also be the case that there are other resistance anisotropies that swamp our proposed contribution or that magnetic breakdown and related subtleties of DSMs could be changing the nature of the cyclotron orbits. We discuss some of these issues in the Appendix \ref{app:DSM}.

A nearly identical calculation for Na$_3$Bi, which has\cite{Liu2014} $k_W \approx 0.095 \AA^{-1}$, yields
\begin{equation}
A_{FS} \sim \frac{e}{\hbar}\left(4 \text{ T}\right)\left(1\pm\frac{\mu}{40 \text{ meV}} \right) \left(\frac{L_{y}}{1 \text{ nm}}\right) \left(\frac{B_{z}}{1 \text{ T}}\right)
\end{equation}

For TaAs, there are twelve pairs of Weyl nodes with varying lengths and curvatures. In particular, some of the Fermi arcs in TaAs have large curvatures and enclose fairly large areas, which will lead to frequency offsets which are independent of in-plane field \cite{PotterFermiArcOsc,ZhangWSMQOs}. With this in mind, using approximate parameters \cite{CASTaAs,PrincetonTaAs} (in particular $2k_W = 0.1-0.5 \AA^{-1}$ for various arcs) we find that the field-dependent parts of the Fermi surface areas are of order
\begin{equation}
A_{FS} \sim \frac{e}{\hbar}\left(1-5 \text{T}\right)\left(\frac{L_{[001]}}{1 \text{ nm}}\right) \left(\frac{B_{||}}{1 \text{ T}}\right)
\label{eqn:TaAsFSEst}
\end{equation}
plus appropriate field-independent offsets. Here $B_{||}$ is the in-plane component of the field. As an important application of our results, our proposal is also able to differentiate between different Fermi arc connection schemes \cite{CASTaAs,PrincetonTaAs} through the dependence of the oscillation frequencies on in-plane field angle. See Appendix \ref{app:materials} for estimations of the offsets for different arcs and details of the angular dependence.

\textit{Discussion}: We have argued both qualitatively and numerically that a closed quasi-2D Fermi surface appears in the thin film quantum limit of Weyl and Dirac semimetals. This Fermi surface leads to unusual quantum oscillations where, in some cases, oscillations only occur as a function of field angle, not of field strength at a fixed angle.

There is another unusual feature of this emergent Fermi surface, which is that the electron wavefunctions with different $k_x$ near the Fermi surface are spatially separated in the $y$ dimension. A consequence is that the effective interactions of low energy electrons are highly anisotropic, as they should be exponentially suppressed in the $k_x$ separation of the states involved, but no suppression occurs in $k_z$ separation. Such anisotropic interactions may have interesting consequences in transport properties, such as a strong anisotropy in the $\propto T^2$ term of the low temperature conductivity. We leave investigations of the consequences of this fact to future work.

\begin{acknowledgments}
\textit{Acknowledgements:} We would like to thank Pavan Hosur, Rex Lundgren, Boris Spivak,  and Yi Zhang for helpful discussions. We acknowledge the hospitality of the Kavli Institute for Theoretical Physics, where some of this work took place. DB is supported by the National Science Foundation under Grant No. DGE-114747. XLQ is supported by the National Science Foundation through the grant No. DMR-1151786.
\end{acknowledgments}

\appendix

\section{Small Gaps in Dirac Semimetals}
\label{app:DSM}

In this section, we detail some subtleties that occur when applying our proposal to Dirac semimetals.

In a Weyl semimetal, the zeroth Landau level at each Weyl point is chiral and thus cannot be gapped out in the bulk. In a Dirac semimetal, on the other hand, there are two zeroth Landau levels with opposite chiralities; they may be gapped out in the same way as the zero-field Dirac cones, that is, via symmetry breaking perturbations or finite size effects. We consider the fate of the quantum oscillations discussed in the main text when such a gap appears.

\begin{figure}
\subfigure[ ]{
\includegraphics[width=5cm]{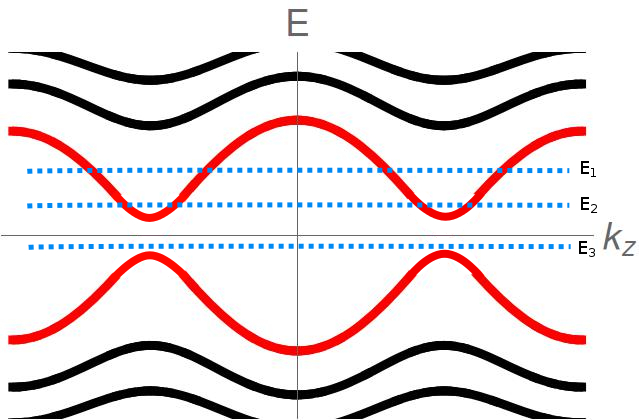}
\label{fig:DiracLLs}
}
\subfigure[ ]{
\includegraphics[width=3cm]{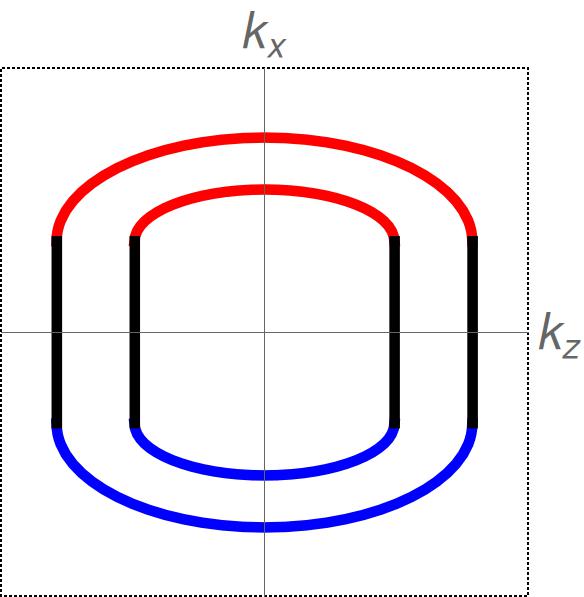}
\label{fig:FSE1}
}
\subfigure[ ]{
\includegraphics[width=3cm]{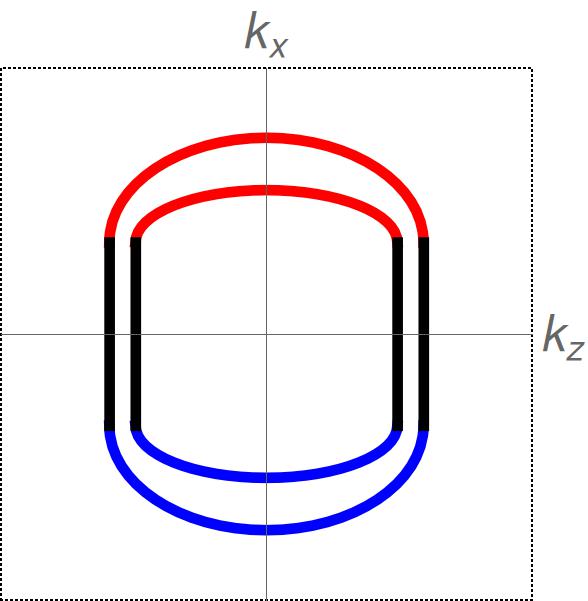}
\label{fig:FSE2}
}
\subfigure[ ]{
\includegraphics[width=3cm]{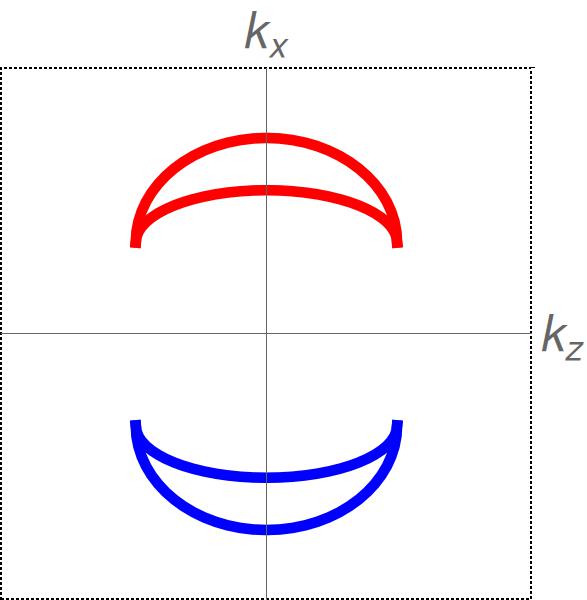}
\label{fig:FSE3}
}
\caption{(a) Schematic bulk Landau level structure of a Dirac semimetal which has been gapped by a symmetry-breaking perturbation or finite size effects. (b)-(d) Schematic Fermi surfaces when the chemical potential is at (b) $E_1$, (c) $E_2$, and (d) $E_3$. In (b)-(d), black states are in the bulk, red states are on the upper surface ,and blue states on the lower surface. (b) is unaffected by the presence of the far-away gap. In (c), the black portions can be close enough so that there is magnetic breakdown during quantum oscillations.}
\end{figure}

Let the gap be $V$. The bulk Landau level structure is shown schematically in Fig. \ref{fig:DiracLLs}. Consider the case where the chemical potential $\mu$ is far from the gap, represented by $\mu = E_1$ in Fig. \ref{fig:DiracLLs}. Then the gap does not affect the states at the chemical potential; looking only at energies close to the chemical potential, we still have two linearly dispersing modes near each Dirac point. Likewise there is no reason for the surface states at the chemical potential to be affected. Hence the gap does not affect the emergent 2D Fermi surfaces, and the Fermi surface looks like Fig. \ref{fig:FSE1}. The same picture would hold even if $V=0$.

If we now allow $\mu$ to approach the gap, say at $\mu = E_2$, the two Fermi surfaces approach each other and approach degeneracy in the bulk, as in Fig. \ref{fig:FSE2}. Once the separation of the bulk portions of the Fermi surface reaches $l_B^{-1}$, magnetic breakdown occurs, allowing electrons to scatter from one Fermi arc to the other one on the same surfce. This suppresses the amplitude of quantum oscillations corresponding to the full Fermi surface. This scattering gets stronger as $\mu$ decreases further, destroying the bulk portion of the oscillations entirely by the time $\mu$ reaches the gap. If we then move $\mu$ into the gap ($\mu = E_3$), the bulk portions (black in Figs. \ref{fig:FSE1} and \ref{fig:FSE2}) of the Fermi surfaces disappear, leaving only the two Fermi arcs on each surface, as shown in Fig. \ref{fig:FSE3}. These Fermi arcs form closed Fermi surfaces whose sizes are insensitive to the magnetic field, and the main result of our paper no longer applies. Instead, the quantum oscillations come purely from surface states and probe the area enclosed by the two Fermi arcs on the same surface.

\section{Material Estimations}
\label{app:materials}

In this section, we elaborate on our estimations for the behavior of quantum oscillations explained in the main text for the Dirac semimetals Cd$_3$As$_2$ and Na$_3$Bi as well as for the 24-node Weyl semimetal TaAs. We discuss constraints on sample thicknesses for all the materials and the highly nontrivial angular dependence of quantum oscillations in TaAs.

\textit{Cd$_3$As$_2$}: We first point out explicitly that the estimations in the main text have assumed that the Fermi arcs in Cd$_3$As$_2$ have no curvature.

To estimate where the ``thick" limit sets in due to finite temperature, we set the temperature scale $\delta \varepsilon/k_B$ from Eq. (9) of the main text to $1$ K at $B_{y} \sim 1$ T. For Cd$_3$As$_2$, assuming $B_{z} \sim 15$ T, we can go as large as $L_y \sim 500$ nm. However, the lifetime constraint Eq. (6) of the main text may cause problems; using the same parameters we find
\begin{equation}
\left(\frac{1 \text{ T}}{B_y}\right)\left(\left(30 \text{fs} \right)\left(\frac{L_y}{1\text{ nm}}\right) + \left(8 \text{ ps}\right)\right) \ll \tau
\label{eqn:Cd3As2lifetime}
\end{equation}
It is unclear what lifetime to compare to; this matters because Ref. \onlinecite{LiangCd3As2} found a discrepancy of four orders of magnitude between the bulk transport lifetime ($\sim$ 500 ps) and the bulk quantum lifetime (30-80 fs). Using the transport lifetime we find a weaker constraint $L_y < 16 \mu$m at $B_y = 1$ T, but using the quantum lifetime no oscillations can occur at all, even at moderately larger values of $B_y$. This latter problem comes from the 8 ps contribution in Eq. (\ref{eqn:Cd3As2lifetime}), which is the time it takes to traverse the Fermi arcs at $B_y = 1$ T. 

Of course, not only are the surface states two-dimensional, but some scattering of the surface states is suppressed (since the other Fermi arc is spatially separated). Hence there is no reason to expect that a bulk lifetime measurement is appropriate for the above estimation. In particular, if we naively use the bulk quantum lifetime in Eq. (\ref{eqn:Cd3As2lifetime}), the 8 ps constraint (the Fermi arc contribution) precludes quantum oscillations even at $B_z=0$, but there is evidence that Fermi arc-induced quantum oscillations have been observed at $B_z=0$ \cite{MollCd3As2Osc}. Furthermore, it has been argued\cite{ZhangWSMQOs} that disorder may play a weaker role than expected even in the bulk. As such, our critical thickness estimates cannot be made more precise until we have a more thorough experimental understanding of how the surfaces and bulk each contribute to the loss of coherence. However, given the experimental history, we are optimistic that our proposed quantum oscillations would appear for reasonable thicknesses.

\textit{Na$_3$Bi:} 
The calculations proceed exactly as for Cd$_3$As$_2$, again assuming no curvature in the Fermi arcs. The Fermi velocity is rather anisotropic; along the Weyl node separation direction $z$ is about $3 \times 10^4$ m/s, almost two orders of magnitude smaller than that of Cd$_3$As$_2$, and the in the other directions it is about $4 \times 10^5$ m/s. The temperature limitation is then, using the same parameters as for Cd$_3$As$_2$, yields $L_y \lesssim 40$ nm. The primary reason for the reduction compared to Cd$_3$As$_2$ is the reduced ratio of Fermi velocities $v_z/v_x$, which is of order 1 in Cd$_3$As$_2$ but is about a tenth in Na$_3$Bi. The lifetime constraint is
\begin{equation}
\left(\frac{1 \text{ T}}{B_y}\right)\left[(1 \text{ ps})\left(\frac{L_y}{1 \text{ nm}}\right) + (22 \text{ ps})\right] \ll \tau
\end{equation}
We are not aware of any quantum lifetime measurements on Na$_3$Bi at present, so we have little to compare to, but expect similar concerns to those in Cd$_3$As$_2$.

\begin{figure}
\subfigure[ ]{
\includegraphics[width=4cm]{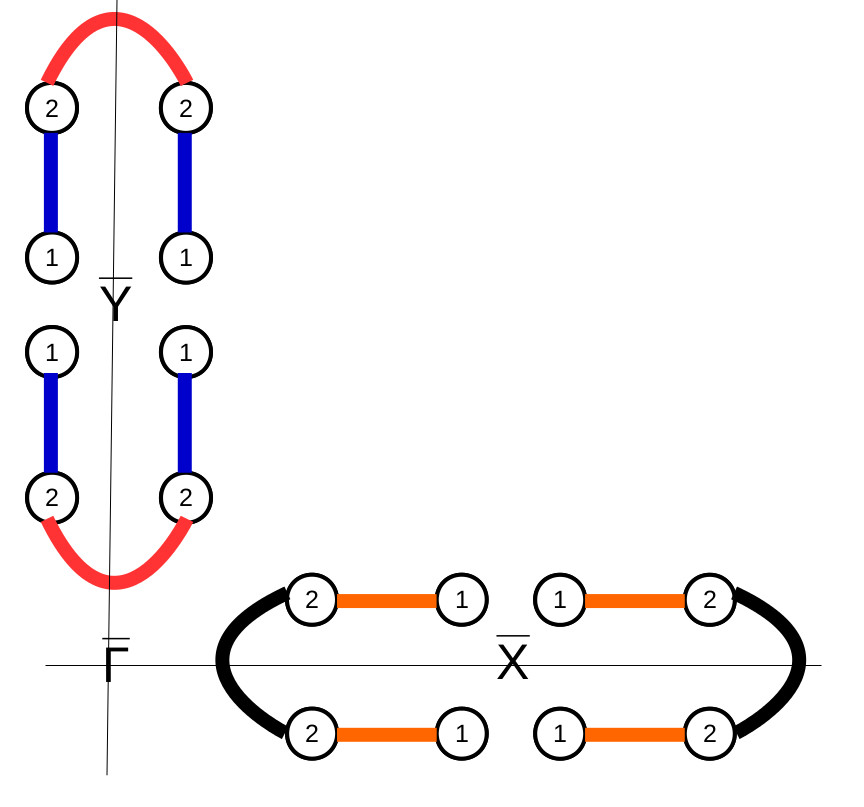}
\label{fig:CASWiring}
}
\subfigure[ ]{
\includegraphics[width=4cm]{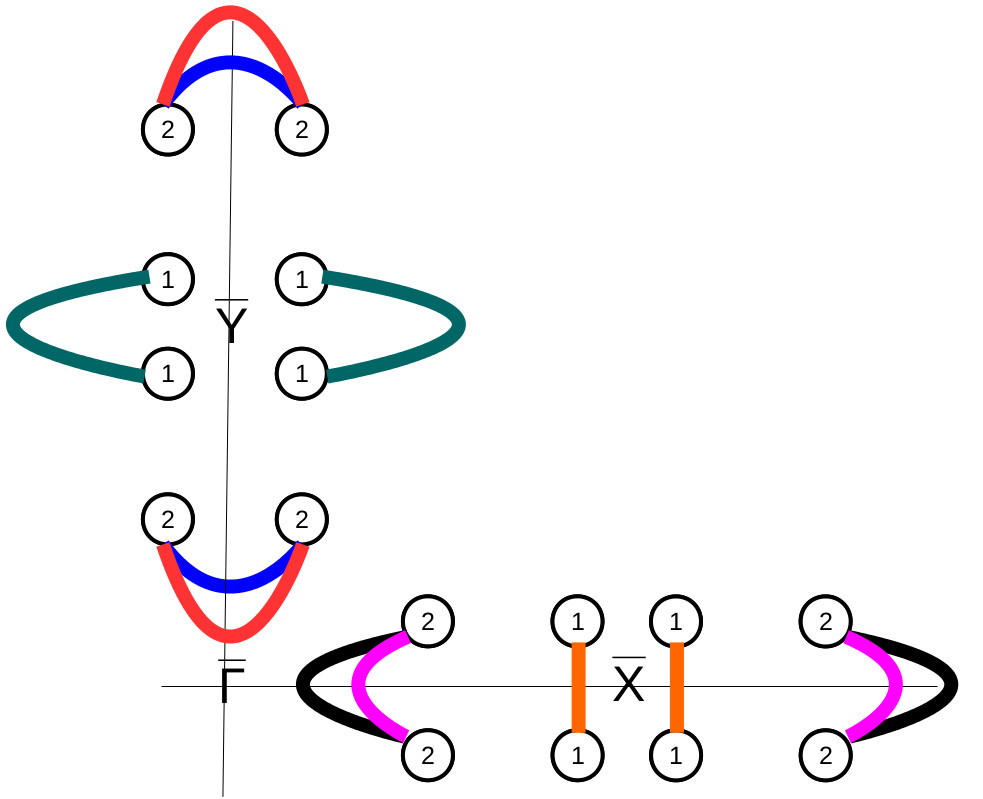}
\label{fig:PrincetonWiring}
}
\subfigure[ ]{
\includegraphics[width=4cm]{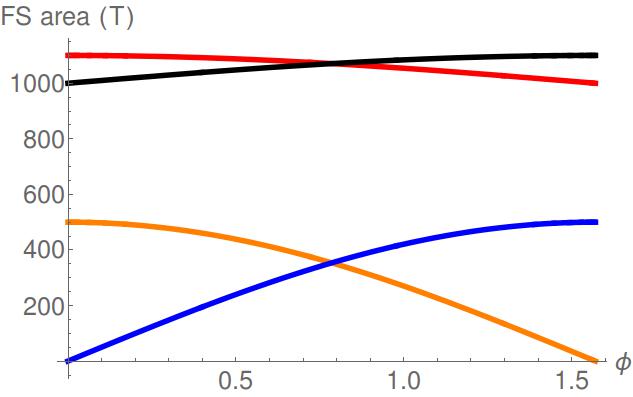}
\label{fig:CASAngle}
}
\subfigure[ ]{
\includegraphics[width=4cm]{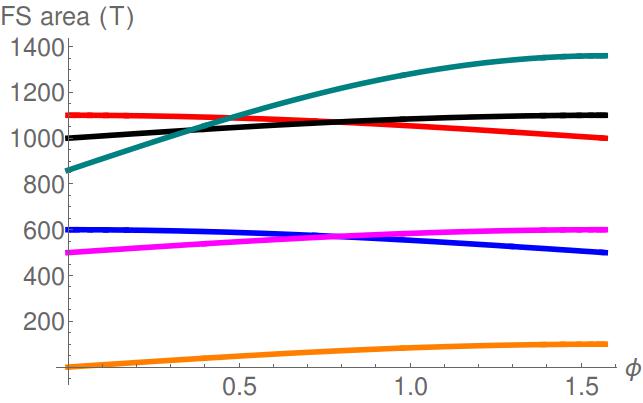}
\label{fig:PrincetonAngle}
}
\caption{(a) and (b): Two possible schematic Fermi arc wiring schemes in TaAs, proposed by (a) Ref. \onlinecite{CASTaAs} and (b) Ref. \onlinecite{PrincetonTaAs}. Circles represent projections of the Weyl nodes onto one surface Brillouin zone, and the number inside is the magnitude of the topological charge of the Weyl node. (c) and (d): Corresponding Fermi surface areas as a function of in-plane field angle $\phi$ for the wiring schemes in (a) and (b) respectively with $B_y = 5$ T and $L_y = 20$ nm. The arcs of a particular color in (a) and (b) produce oscillation frequencies with the same color in (c) and (d) respectively.}
\label{fig:wirings}
\end{figure}

\textit{TaAs}: Our estimations for TaAs have several caveats. First, different \textit{ab initio} calculations disagree on which Weyl points are connected by Fermi arcs, and ARPES does not currently have sufficient resolution to decide which bands are Fermi arcs and which are trivial surface bands. In addition to quantitative effects, this qualitatively affects how many differently sized Fermi dependences on in-plane field angle. Conveniently, our proposal predicts that the in-plane angle dependence of these emergent Fermi surfaces has the same symmetry as the Fermi arc configuration, so measurements of our proposal could fairly easily distinguish $C_4$-symmetric Fermi arc configurations \cite{CASTaAs} as shown in Fig. \ref{fig:CASWiring} from asymmetric ones \cite{PrincetonTaAs} as shown in Fig. \ref{fig:PrincetonWiring}. Second, some parallel Fermi arcs are quite close to each other; presumably if the corresponding 2D Fermi surfaces overlap with each other, there can be hybridization (depending on the magnetic length).

We first estimate where the thick-film limit occurs. Under the same estimations as for Cd$_3$As$_2$, the temperature-enforced thick limit sets in at $L_y \sim 100$ nm. (Typical crystals are cleaved in the $[001]$ plane; we adopt $[001]$ as the $y$ direction for consistency with our previous notation.) The scattering constraint yields
\begin{equation}
\left(\frac{1 \text{ T}}{B_y}\right)\left(\left(0.3 \text{ ps} \right)\left(\frac{L_y}{1\text{ nm}}\right) + \left(20 \text{ ps}\right)\right) \ll \tau
\label{eqn:TaAslifetime}
\end{equation}
Ref. \onlinecite{ZhangTaAsTransport} measured a bulk quantum lifetime of $0.4$ ps, but for reasons similar to that of Cd$_3$As$_2$, more experimental work is needed to determing the appropriate lifetime to use in this constraint.

The in-plane angular dependence, as mentioned previously, depends on the arc configuration. Recall that Eq. (4) of the main text tells us that the frequency depends on the component of the field along the splitting of the Weyl points connected by Fermi arcs. In the scheme in Fig. \ref{fig:CASWiring}, there are two types of Fermi arc up to $C_4$ rotational symmetry. One type contains no area, and thus leads to one frequency which goes as $|\cos \phi|$ (blue lines) and one that goes as $|\sin \phi|$ (orange lines). The other type has Fermi arcs which enclose an area of about $1$ kT, so those will lead to two frequencies centered at $\sim 1$ kT but offset by the frequency in Eq. (11) of the main text (about 1-5 $T$) times $|\cos \phi|$ (red arcs) and $|\sin \phi|$ (blue arcs) respectively. In total there are four different frequencies; their angular dependences are plotted in Fig. \ref{fig:CASAngle}.

In the $C_4$-asymmetric scheme in Fig. \ref{fig:PrincetonWiring}, there are a total of six different frequencies. One frequency goes as $|\sin \phi|$ (orange lines). The other Fermi arcs have curvature and thus have offsets. Three go as $|\sin \phi|$ and have offsets of order $1$ kT; these offsets differ by some factor of order 1 (we took ballpark estimates of $500$ T, $800$ T, and $1$ kT (magenta, blue-green, and black arcs respectively) based on the experimental data in Ref. \onlinecite{PrincetonTaAs}). The remaining arcs go as $|\cos \phi|$ with offsets $\sim 500$ T and $\sim 1$ kT (blue and red arcs respectively). The angular dependences are plotted in Fig. \ref{fig:PrincetonAngle}.

\bibstyle{apsrev4-1} \bibliography{references}

\begin{thebibliography}{34}%
\makeatletter
\providecommand \@ifxundefined [1]{%
 \@ifx{#1\undefined}
}%
\providecommand \@ifnum [1]{%
 \ifnum #1\expandafter \@firstoftwo
 \else \expandafter \@secondoftwo
 \fi
}%
\providecommand \@ifx [1]{%
 \ifx #1\expandafter \@firstoftwo
 \else \expandafter \@secondoftwo
 \fi
}%
\providecommand \natexlab [1]{#1}%
\providecommand \enquote  [1]{``#1''}%
\providecommand \bibnamefont  [1]{#1}%
\providecommand \bibfnamefont [1]{#1}%
\providecommand \citenamefont [1]{#1}%
\providecommand \href@noop [0]{\@secondoftwo}%
\providecommand \href [0]{\begingroup \@sanitize@url \@href}%
\providecommand \@href[1]{\@@startlink{#1}\@@href}%
\providecommand \@@href[1]{\endgroup#1\@@endlink}%
\providecommand \@sanitize@url [0]{\catcode `\\12\catcode `\$12\catcode
  `\&12\catcode `\#12\catcode `\^12\catcode `\_12\catcode `\%12\relax}%
\providecommand \@@startlink[1]{}%
\providecommand \@@endlink[0]{}%
\providecommand \url  [0]{\begingroup\@sanitize@url \@url }%
\providecommand \@url [1]{\endgroup\@href {#1}{\urlprefix }}%
\providecommand \urlprefix  [0]{URL }%
\providecommand \Eprint [0]{\href }%
\providecommand \doibase [0]{http://dx.doi.org/}%
\providecommand \selectlanguage [0]{\@gobble}%
\providecommand \bibinfo  [0]{\@secondoftwo}%
\providecommand \bibfield  [0]{\@secondoftwo}%
\providecommand \translation [1]{[#1]}%
\providecommand \BibitemOpen [0]{}%
\providecommand \bibitemStop [0]{}%
\providecommand \bibitemNoStop [0]{.\EOS\space}%
\providecommand \EOS [0]{\spacefactor3000\relax}%
\providecommand \BibitemShut  [1]{\csname bibitem#1\endcsname}%
\let\auto@bib@innerbib\@empty
\bibitem [{\citenamefont {Murakami}(2007)}]{Murakami2007}%
  \BibitemOpen
  \bibfield  {author} {\bibinfo {author} {\bibfnamefont {S.}~\bibnamefont
  {Murakami}},\ }\href {http://stacks.iop.org/1367-2630/9/i=9/a=356} {\bibfield
   {journal} {\bibinfo  {journal} {New J. of Phys.}\ }\textbf {\bibinfo
  {volume} {9}},\ \bibinfo {pages} {356} (\bibinfo {year} {2007})}\BibitemShut
  {NoStop}%
\bibitem [{\citenamefont {Wan}\ \emph {et~al.}(2011)\citenamefont {Wan},
  \citenamefont {Turner}, \citenamefont {Vishwanath},\ and\ \citenamefont
  {Savrasov}}]{PyrochloreWeyl}%
  \BibitemOpen
  \bibfield  {author} {\bibinfo {author} {\bibfnamefont {X.}~\bibnamefont
  {Wan}}, \bibinfo {author} {\bibfnamefont {A.~M.}\ \bibnamefont {Turner}},
  \bibinfo {author} {\bibfnamefont {A.}~\bibnamefont {Vishwanath}}, \ and\
  \bibinfo {author} {\bibfnamefont {S.~Y.}\ \bibnamefont {Savrasov}},\ }\href
  {\doibase 10.1103/PhysRevB.83.205101} {\bibfield  {journal} {\bibinfo
  {journal} {Phys. Rev. B}\ }\textbf {\bibinfo {volume} {83}},\ \bibinfo
  {pages} {205101} (\bibinfo {year} {2011})}\BibitemShut {NoStop}%
\bibitem [{\citenamefont {Burkov}\ and\ \citenamefont
  {Balents}(2011)}]{WeylMultiLayer}%
  \BibitemOpen
  \bibfield  {author} {\bibinfo {author} {\bibfnamefont {A.~A.}\ \bibnamefont
  {Burkov}}\ and\ \bibinfo {author} {\bibfnamefont {L.}~\bibnamefont
  {Balents}},\ }\href {\doibase 10.1103/PhysRevLett.107.127205} {\bibfield
  {journal} {\bibinfo  {journal} {Phys. Rev. Lett.}\ }\textbf {\bibinfo
  {volume} {107}},\ \bibinfo {pages} {127205} (\bibinfo {year}
  {2011})}\BibitemShut {NoStop}%
\bibitem [{\citenamefont {Young}\ \emph {et~al.}(2012)\citenamefont {Young},
  \citenamefont {Zaheer}, \citenamefont {Teo}, \citenamefont {Kane},
  \citenamefont {Mele},\ and\ \citenamefont {Rappe}}]{YoungDiracSemimetal}%
  \BibitemOpen
  \bibfield  {author} {\bibinfo {author} {\bibfnamefont {S.~M.}\ \bibnamefont
  {Young}}, \bibinfo {author} {\bibfnamefont {S.}~\bibnamefont {Zaheer}},
  \bibinfo {author} {\bibfnamefont {J.~C.~Y.}\ \bibnamefont {Teo}}, \bibinfo
  {author} {\bibfnamefont {C.~L.}\ \bibnamefont {Kane}}, \bibinfo {author}
  {\bibfnamefont {E.~J.}\ \bibnamefont {Mele}}, \ and\ \bibinfo {author}
  {\bibfnamefont {A.~M.}\ \bibnamefont {Rappe}},\ }\href {\doibase
  10.1103/PhysRevLett.108.140405} {\bibfield  {journal} {\bibinfo  {journal}
  {Phys. Rev. Lett.}\ }\textbf {\bibinfo {volume} {108}},\ \bibinfo {pages}
  {140405} (\bibinfo {year} {2012})}\BibitemShut {NoStop}%
\bibitem [{\citenamefont {Zyuzin}\ and\ \citenamefont
  {Burkov}(2012)}]{ZyuninBurkovWeylTheta}%
  \BibitemOpen
  \bibfield  {author} {\bibinfo {author} {\bibfnamefont {A.~A.}\ \bibnamefont
  {Zyuzin}}\ and\ \bibinfo {author} {\bibfnamefont {A.~A.}\ \bibnamefont
  {Burkov}},\ }\href {\doibase 10.1103/PhysRevB.86.115133} {\bibfield
  {journal} {\bibinfo  {journal} {Phys. Rev. B}\ }\textbf {\bibinfo {volume}
  {86}},\ \bibinfo {pages} {115133} (\bibinfo {year} {2012})}\BibitemShut
  {NoStop}%
\bibitem [{\citenamefont {Son}\ and\ \citenamefont
  {Spivak}(2013)}]{SonSpivakWeylAnomaly}%
  \BibitemOpen
  \bibfield  {author} {\bibinfo {author} {\bibfnamefont {D.~T.}\ \bibnamefont
  {Son}}\ and\ \bibinfo {author} {\bibfnamefont {B.~Z.}\ \bibnamefont
  {Spivak}},\ }\href {\doibase 10.1103/PhysRevB.88.104412} {\bibfield
  {journal} {\bibinfo  {journal} {Phys. Rev. B}\ }\textbf {\bibinfo {volume}
  {88}},\ \bibinfo {pages} {104412} (\bibinfo {year} {2013})}\BibitemShut
  {NoStop}%
\bibitem [{\citenamefont {Liu}\ \emph {et~al.}(2013)\citenamefont {Liu},
  \citenamefont {Ye},\ and\ \citenamefont {Qi}}]{QiWeylAnomaly}%
  \BibitemOpen
  \bibfield  {author} {\bibinfo {author} {\bibfnamefont {C.-X.}\ \bibnamefont
  {Liu}}, \bibinfo {author} {\bibfnamefont {P.}~\bibnamefont {Ye}}, \ and\
  \bibinfo {author} {\bibfnamefont {X.-L.}\ \bibnamefont {Qi}},\ }\href
  {\doibase 10.1103/PhysRevB.87.235306} {\bibfield  {journal} {\bibinfo
  {journal} {Phys. Rev. B}\ }\textbf {\bibinfo {volume} {87}},\ \bibinfo
  {pages} {235306} (\bibinfo {year} {2013})}\BibitemShut {NoStop}%
\bibitem [{\citenamefont {Hosur}\ and\ \citenamefont
  {Qi}(2013)}]{HosurWeylReview}%
  \BibitemOpen
  \bibfield  {author} {\bibinfo {author} {\bibfnamefont {P.}~\bibnamefont
  {Hosur}}\ and\ \bibinfo {author} {\bibfnamefont {X.}~\bibnamefont {Qi}},\
  }\bibfield  {booktitle} {\emph {\bibinfo {booktitle} {Topological insulators
  / Isolants topologiques}},\ }\href
  {http://www.sciencedirect.com/science/article/pii/S1631070513001710}
  {\bibfield  {journal} {\bibinfo  {journal} {Comptes Rendus Physique}\
  }\textbf {\bibinfo {volume} {14}},\ \bibinfo {pages} {857} (\bibinfo {year}
  {2013})}\BibitemShut {NoStop}%
\bibitem [{\citenamefont {Wang}\ \emph {et~al.}(2012)\citenamefont {Wang},
  \citenamefont {Sun}, \citenamefont {Chen}, \citenamefont {Franchini},
  \citenamefont {Xu}, \citenamefont {Weng}, \citenamefont {Dai},\ and\
  \citenamefont {Fang}}]{WangA3Bi}%
  \BibitemOpen
  \bibfield  {author} {\bibinfo {author} {\bibfnamefont {Z.}~\bibnamefont
  {Wang}}, \bibinfo {author} {\bibfnamefont {Y.}~\bibnamefont {Sun}}, \bibinfo
  {author} {\bibfnamefont {X.-Q.}\ \bibnamefont {Chen}}, \bibinfo {author}
  {\bibfnamefont {C.}~\bibnamefont {Franchini}}, \bibinfo {author}
  {\bibfnamefont {G.}~\bibnamefont {Xu}}, \bibinfo {author} {\bibfnamefont
  {H.}~\bibnamefont {Weng}}, \bibinfo {author} {\bibfnamefont {X.}~\bibnamefont
  {Dai}}, \ and\ \bibinfo {author} {\bibfnamefont {Z.}~\bibnamefont {Fang}},\
  }\href {\doibase 10.1103/PhysRevB.85.195320} {\bibfield  {journal} {\bibinfo
  {journal} {Phys. Rev. B}\ }\textbf {\bibinfo {volume} {85}},\ \bibinfo
  {pages} {195320} (\bibinfo {year} {2012})}\BibitemShut {NoStop}%
\bibitem [{\citenamefont {Xu}\ \emph {et~al.}(2013)\citenamefont {Xu},
  \citenamefont {Liu}, \citenamefont {Kushwaha}, \citenamefont {Chang},
  \citenamefont {Krizan}, \citenamefont {Sankar}, \citenamefont {Polley},
  \citenamefont {Adell}, \citenamefont {Balasubramanian}, \citenamefont
  {Miyamoto}, \citenamefont {Alidoust}, \citenamefont {Bian}, \citenamefont
  {Neupane}, \citenamefont {Belopolski}, \citenamefont {Jeng}, \citenamefont
  {Huang}, \citenamefont {Tsai}, \citenamefont {Chou}, \citenamefont {Okuda},
  \citenamefont {Bansil}, \citenamefont {Cava},\ and\ \citenamefont
  {Hasan}}]{Xu2013}%
  \BibitemOpen
  \bibfield  {author} {\bibinfo {author} {\bibfnamefont {S.-Y.}\ \bibnamefont
  {Xu}}, \bibinfo {author} {\bibfnamefont {C.}~\bibnamefont {Liu}}, \bibinfo
  {author} {\bibfnamefont {S.~K.}\ \bibnamefont {Kushwaha}}, \bibinfo {author}
  {\bibfnamefont {T.-R.}\ \bibnamefont {Chang}}, \bibinfo {author}
  {\bibfnamefont {J.~W.}\ \bibnamefont {Krizan}}, \bibinfo {author}
  {\bibfnamefont {R.}~\bibnamefont {Sankar}}, \bibinfo {author} {\bibfnamefont
  {C.~M.}\ \bibnamefont {Polley}}, \bibinfo {author} {\bibfnamefont
  {J.}~\bibnamefont {Adell}}, \bibinfo {author} {\bibfnamefont
  {T.}~\bibnamefont {Balasubramanian}}, \bibinfo {author} {\bibfnamefont
  {K.}~\bibnamefont {Miyamoto}}, \bibinfo {author} {\bibfnamefont
  {N.}~\bibnamefont {Alidoust}}, \bibinfo {author} {\bibfnamefont
  {G.}~\bibnamefont {Bian}}, \bibinfo {author} {\bibfnamefont {M.}~\bibnamefont
  {Neupane}}, \bibinfo {author} {\bibfnamefont {I.}~\bibnamefont {Belopolski}},
  \bibinfo {author} {\bibfnamefont {H.-T.}\ \bibnamefont {Jeng}}, \bibinfo
  {author} {\bibfnamefont {C.-Y.}\ \bibnamefont {Huang}}, \bibinfo {author}
  {\bibfnamefont {W.-F.}\ \bibnamefont {Tsai}}, \bibinfo {author}
  {\bibfnamefont {H.~L. F.~C.}\ \bibnamefont {Chou}}, \bibinfo {author}
  {\bibfnamefont {T.}~\bibnamefont {Okuda}}, \bibinfo {author} {\bibfnamefont
  {A.}~\bibnamefont {Bansil}}, \bibinfo {author} {\bibfnamefont {R.~J.}\
  \bibnamefont {Cava}}, \ and\ \bibinfo {author} {\bibfnamefont {M.~Z.}\
  \bibnamefont {Hasan}},\ }\href@noop {} {\bibfield  {journal} {\bibinfo
  {journal} {arXiv e-prints}\ } (\bibinfo {year} {2013})},\ \Eprint
  {http://arxiv.org/abs/1312.7624} {arXiv:1312.7624} \BibitemShut {NoStop}%
\bibitem [{\citenamefont {Liu}\ \emph {et~al.}(2014{\natexlab{a}})\citenamefont
  {Liu}, \citenamefont {Zhou}, \citenamefont {Zhang}, \citenamefont {Wang},
  \citenamefont {Weng}, \citenamefont {Prabhakaran}, \citenamefont {Mo},
  \citenamefont {Shen}, \citenamefont {Fang}, \citenamefont {Dai},
  \citenamefont {Hussain},\ and\ \citenamefont {Chen}}]{Liu2014}%
  \BibitemOpen
  \bibfield  {author} {\bibinfo {author} {\bibfnamefont {Z.~K.}\ \bibnamefont
  {Liu}}, \bibinfo {author} {\bibfnamefont {B.}~\bibnamefont {Zhou}}, \bibinfo
  {author} {\bibfnamefont {Y.}~\bibnamefont {Zhang}}, \bibinfo {author}
  {\bibfnamefont {Z.~J.}\ \bibnamefont {Wang}}, \bibinfo {author}
  {\bibfnamefont {H.~M.}\ \bibnamefont {Weng}}, \bibinfo {author}
  {\bibfnamefont {D.}~\bibnamefont {Prabhakaran}}, \bibinfo {author}
  {\bibfnamefont {S.-K.}\ \bibnamefont {Mo}}, \bibinfo {author} {\bibfnamefont
  {Z.~X.}\ \bibnamefont {Shen}}, \bibinfo {author} {\bibfnamefont
  {Z.}~\bibnamefont {Fang}}, \bibinfo {author} {\bibfnamefont {X.}~\bibnamefont
  {Dai}}, \bibinfo {author} {\bibfnamefont {Z.}~\bibnamefont {Hussain}}, \ and\
  \bibinfo {author} {\bibfnamefont {Y.~L.}\ \bibnamefont {Chen}},\ }\href
  {\doibase 10.1126/science.1245085} {\bibfield  {journal} {\bibinfo  {journal}
  {Science}\ }\textbf {\bibinfo {volume} {343}},\ \bibinfo {pages} {864}
  (\bibinfo {year} {2014}{\natexlab{a}})}\BibitemShut {NoStop}%
\bibitem [{\citenamefont {Wang}\ \emph {et~al.}(2013)\citenamefont {Wang},
  \citenamefont {Weng}, \citenamefont {Wu}, \citenamefont {Dai},\ and\
  \citenamefont {Fang}}]{WangCd3As2}%
  \BibitemOpen
  \bibfield  {author} {\bibinfo {author} {\bibfnamefont {Z.}~\bibnamefont
  {Wang}}, \bibinfo {author} {\bibfnamefont {H.}~\bibnamefont {Weng}}, \bibinfo
  {author} {\bibfnamefont {Q.}~\bibnamefont {Wu}}, \bibinfo {author}
  {\bibfnamefont {X.}~\bibnamefont {Dai}}, \ and\ \bibinfo {author}
  {\bibfnamefont {Z.}~\bibnamefont {Fang}},\ }\href {\doibase
  10.1103/PhysRevB.88.125427} {\bibfield  {journal} {\bibinfo  {journal} {Phys.
  Rev. B}\ }\textbf {\bibinfo {volume} {88}},\ \bibinfo {pages} {125427}
  (\bibinfo {year} {2013})}\BibitemShut {NoStop}%
\bibitem [{\citenamefont {Liu}\ \emph {et~al.}(2014{\natexlab{b}})\citenamefont
  {Liu}, \citenamefont {Jiang}, \citenamefont {Zhou}, \citenamefont {Wang},
  \citenamefont {Zhang}, \citenamefont {Weng}, \citenamefont {Prabhakaran},
  \citenamefont {Mo}, \citenamefont {Peng}, \citenamefont {Dudin},
  \citenamefont {Kim}, \citenamefont {Hoesch}, \citenamefont {Fang},
  \citenamefont {Dai}, \citenamefont {Shen}, \citenamefont {Feng},
  \citenamefont {Hussain},\ and\ \citenamefont {Chen}}]{LiuCd3As2}%
  \BibitemOpen
  \bibfield  {author} {\bibinfo {author} {\bibfnamefont {Z.~K.}\ \bibnamefont
  {Liu}}, \bibinfo {author} {\bibfnamefont {J.}~\bibnamefont {Jiang}}, \bibinfo
  {author} {\bibfnamefont {B.}~\bibnamefont {Zhou}}, \bibinfo {author}
  {\bibfnamefont {Z.~J.}\ \bibnamefont {Wang}}, \bibinfo {author}
  {\bibfnamefont {Y.}~\bibnamefont {Zhang}}, \bibinfo {author} {\bibfnamefont
  {H.~M.}\ \bibnamefont {Weng}}, \bibinfo {author} {\bibfnamefont
  {D.}~\bibnamefont {Prabhakaran}}, \bibinfo {author} {\bibfnamefont {S.-K.}\
  \bibnamefont {Mo}}, \bibinfo {author} {\bibfnamefont {H.}~\bibnamefont
  {Peng}}, \bibinfo {author} {\bibfnamefont {P.}~\bibnamefont {Dudin}},
  \bibinfo {author} {\bibfnamefont {T.}~\bibnamefont {Kim}}, \bibinfo {author}
  {\bibfnamefont {M.}~\bibnamefont {Hoesch}}, \bibinfo {author} {\bibfnamefont
  {Z.}~\bibnamefont {Fang}}, \bibinfo {author} {\bibfnamefont {X.}~\bibnamefont
  {Dai}}, \bibinfo {author} {\bibfnamefont {Z.~X.}\ \bibnamefont {Shen}},
  \bibinfo {author} {\bibfnamefont {D.~L.}\ \bibnamefont {Feng}}, \bibinfo
  {author} {\bibfnamefont {Z.}~\bibnamefont {Hussain}}, \ and\ \bibinfo
  {author} {\bibfnamefont {Y.~L.}\ \bibnamefont {Chen}},\ }\href {\doibase
  10.1038/nmat3990} {\bibfield  {journal} {\bibinfo  {journal} {Nat. Mater.}\
  }\textbf {\bibinfo {volume} {13}},\ \bibinfo {pages} {677} (\bibinfo {year}
  {2014}{\natexlab{b}})}\BibitemShut {NoStop}%
\bibitem [{\citenamefont {Neupane}\ \emph {et~al.}(2014)\citenamefont
  {Neupane}, \citenamefont {Xu}, \citenamefont {Sankar}, \citenamefont
  {Alidoust}, \citenamefont {Bian}, \citenamefont {Liu}, \citenamefont
  {Belopolski}, \citenamefont {Chang}, \citenamefont {Jeng}, \citenamefont
  {Lin}, \citenamefont {Bansil}, \citenamefont {Chou},\ and\ \citenamefont
  {Hasan}}]{Neupane2014}%
  \BibitemOpen
  \bibfield  {author} {\bibinfo {author} {\bibfnamefont {M.}~\bibnamefont
  {Neupane}}, \bibinfo {author} {\bibfnamefont {S.-Y.}\ \bibnamefont {Xu}},
  \bibinfo {author} {\bibfnamefont {R.}~\bibnamefont {Sankar}}, \bibinfo
  {author} {\bibfnamefont {N.}~\bibnamefont {Alidoust}}, \bibinfo {author}
  {\bibfnamefont {G.}~\bibnamefont {Bian}}, \bibinfo {author} {\bibfnamefont
  {C.}~\bibnamefont {Liu}}, \bibinfo {author} {\bibfnamefont {I.}~\bibnamefont
  {Belopolski}}, \bibinfo {author} {\bibfnamefont {T.-R.}\ \bibnamefont
  {Chang}}, \bibinfo {author} {\bibfnamefont {H.-T.}\ \bibnamefont {Jeng}},
  \bibinfo {author} {\bibfnamefont {H.}~\bibnamefont {Lin}}, \bibinfo {author}
  {\bibfnamefont {A.}~\bibnamefont {Bansil}}, \bibinfo {author} {\bibfnamefont
  {F.}~\bibnamefont {Chou}}, \ and\ \bibinfo {author} {\bibfnamefont {M.~Z.}\
  \bibnamefont {Hasan}},\ }\href {\doibase 10.1038/ncomms4786} {\bibfield
  {journal} {\bibinfo  {journal} {Nat. Commun.}\ }\textbf {\bibinfo {volume}
  {5}},\ \bibinfo {pages} {3786} (\bibinfo {year} {2014})}\BibitemShut
  {NoStop}%
\bibitem [{\citenamefont {Borisenko}\ \emph {et~al.}(2014)\citenamefont
  {Borisenko}, \citenamefont {Gibson}, \citenamefont {Evtushinsky},
  \citenamefont {Zabolotnyy}, \citenamefont {B\"uchner},\ and\ \citenamefont
  {Cava}}]{BorisenkoCd3As2}%
  \BibitemOpen
  \bibfield  {author} {\bibinfo {author} {\bibfnamefont {S.}~\bibnamefont
  {Borisenko}}, \bibinfo {author} {\bibfnamefont {Q.}~\bibnamefont {Gibson}},
  \bibinfo {author} {\bibfnamefont {D.}~\bibnamefont {Evtushinsky}}, \bibinfo
  {author} {\bibfnamefont {V.}~\bibnamefont {Zabolotnyy}}, \bibinfo {author}
  {\bibfnamefont {B.}~\bibnamefont {B\"uchner}}, \ and\ \bibinfo {author}
  {\bibfnamefont {R.~J.}\ \bibnamefont {Cava}},\ }\href {\doibase
  10.1103/PhysRevLett.113.027603} {\bibfield  {journal} {\bibinfo  {journal}
  {Phys. Rev. Lett.}\ }\textbf {\bibinfo {volume} {113}},\ \bibinfo {pages}
  {027603} (\bibinfo {year} {2014})}\BibitemShut {NoStop}%
\bibitem [{\citenamefont {Weng}\ \emph {et~al.}(2015)\citenamefont {Weng},
  \citenamefont {Fang}, \citenamefont {Fang}, \citenamefont {Bernevig},\ and\
  \citenamefont {Dai}}]{TaAsPredictionPRX}%
  \BibitemOpen
  \bibfield  {author} {\bibinfo {author} {\bibfnamefont {H.}~\bibnamefont
  {Weng}}, \bibinfo {author} {\bibfnamefont {C.}~\bibnamefont {Fang}}, \bibinfo
  {author} {\bibfnamefont {Z.}~\bibnamefont {Fang}}, \bibinfo {author}
  {\bibfnamefont {B.~A.}\ \bibnamefont {Bernevig}}, \ and\ \bibinfo {author}
  {\bibfnamefont {X.}~\bibnamefont {Dai}},\ }\href {\doibase
  10.1103/PhysRevX.5.011029} {\bibfield  {journal} {\bibinfo  {journal} {Phys.
  Rev. X}\ }\textbf {\bibinfo {volume} {5}},\ \bibinfo {pages} {011029}
  (\bibinfo {year} {2015})}\BibitemShut {NoStop}%
\bibitem [{\citenamefont {Huang}\ \emph {et~al.}(2015)\citenamefont {Huang},
  \citenamefont {Xu}, \citenamefont {Belopolski}, \citenamefont {Lee},
  \citenamefont {Chang}, \citenamefont {Wang}, \citenamefont {Alidoust},
  \citenamefont {Bian}, \citenamefont {Neupane}, \citenamefont {Zhang},
  \citenamefont {Jia}, \citenamefont {Bansil}, \citenamefont {Lin},\ and\
  \citenamefont {Hasan}}]{TaAsPrediction}%
  \BibitemOpen
  \bibfield  {author} {\bibinfo {author} {\bibfnamefont {S.-M.}\ \bibnamefont
  {Huang}}, \bibinfo {author} {\bibfnamefont {S.-Y.}\ \bibnamefont {Xu}},
  \bibinfo {author} {\bibfnamefont {I.}~\bibnamefont {Belopolski}}, \bibinfo
  {author} {\bibfnamefont {C.-C.}\ \bibnamefont {Lee}}, \bibinfo {author}
  {\bibfnamefont {G.}~\bibnamefont {Chang}}, \bibinfo {author} {\bibfnamefont
  {B.}~\bibnamefont {Wang}}, \bibinfo {author} {\bibfnamefont {N.}~\bibnamefont
  {Alidoust}}, \bibinfo {author} {\bibfnamefont {G.}~\bibnamefont {Bian}},
  \bibinfo {author} {\bibfnamefont {M.}~\bibnamefont {Neupane}}, \bibinfo
  {author} {\bibfnamefont {C.}~\bibnamefont {Zhang}}, \bibinfo {author}
  {\bibfnamefont {S.}~\bibnamefont {Jia}}, \bibinfo {author} {\bibfnamefont
  {A.}~\bibnamefont {Bansil}}, \bibinfo {author} {\bibfnamefont
  {H.}~\bibnamefont {Lin}}, \ and\ \bibinfo {author} {\bibfnamefont {M.~Z.}\
  \bibnamefont {Hasan}},\ }\href {\doibase 10.1038/ncomms8373} {\bibfield
  {journal} {\bibinfo  {journal} {Nat. Commun.}\ }\textbf {\bibinfo {volume}
  {6}},\ \bibinfo {pages} {7373} (\bibinfo {year} {2015})}\BibitemShut
  {NoStop}%
\bibitem [{\citenamefont {Xu}\ \emph {et~al.}(2015)\citenamefont {Xu},
  \citenamefont {Belopolski}, \citenamefont {Alidoust}, \citenamefont
  {Neupane}, \citenamefont {Bian}, \citenamefont {Zhang}, \citenamefont
  {Sankar}, \citenamefont {Chang}, \citenamefont {Yuan}, \citenamefont {Lee},
  \citenamefont {Huang}, \citenamefont {Zheng}, \citenamefont {Ma},
  \citenamefont {Sanchez}, \citenamefont {Wang}, \citenamefont {Bansil},
  \citenamefont {Chou}, \citenamefont {Shibayev}, \citenamefont {Lin},
  \citenamefont {Jia},\ and\ \citenamefont {Hasan}}]{PrincetonTaAs}%
  \BibitemOpen
  \bibfield  {author} {\bibinfo {author} {\bibfnamefont {S.-Y.}\ \bibnamefont
  {Xu}}, \bibinfo {author} {\bibfnamefont {I.}~\bibnamefont {Belopolski}},
  \bibinfo {author} {\bibfnamefont {N.}~\bibnamefont {Alidoust}}, \bibinfo
  {author} {\bibfnamefont {M.}~\bibnamefont {Neupane}}, \bibinfo {author}
  {\bibfnamefont {G.}~\bibnamefont {Bian}}, \bibinfo {author} {\bibfnamefont
  {C.}~\bibnamefont {Zhang}}, \bibinfo {author} {\bibfnamefont
  {R.}~\bibnamefont {Sankar}}, \bibinfo {author} {\bibfnamefont
  {G.}~\bibnamefont {Chang}}, \bibinfo {author} {\bibfnamefont
  {Z.}~\bibnamefont {Yuan}}, \bibinfo {author} {\bibfnamefont {C.-C.}\
  \bibnamefont {Lee}}, \bibinfo {author} {\bibfnamefont {S.-M.}\ \bibnamefont
  {Huang}}, \bibinfo {author} {\bibfnamefont {H.}~\bibnamefont {Zheng}},
  \bibinfo {author} {\bibfnamefont {J.}~\bibnamefont {Ma}}, \bibinfo {author}
  {\bibfnamefont {D.~S.}\ \bibnamefont {Sanchez}}, \bibinfo {author}
  {\bibfnamefont {B.}~\bibnamefont {Wang}}, \bibinfo {author} {\bibfnamefont
  {A.}~\bibnamefont {Bansil}}, \bibinfo {author} {\bibfnamefont
  {F.}~\bibnamefont {Chou}}, \bibinfo {author} {\bibfnamefont {P.~P.}\
  \bibnamefont {Shibayev}}, \bibinfo {author} {\bibfnamefont {H.}~\bibnamefont
  {Lin}}, \bibinfo {author} {\bibfnamefont {S.}~\bibnamefont {Jia}}, \ and\
  \bibinfo {author} {\bibfnamefont {M.~Z.}\ \bibnamefont {Hasan}},\ }\href
  {\doibase 10.1126/science.aaa9297} {\bibfield  {journal} {\bibinfo  {journal}
  {Science}\ }\textbf {\bibinfo {volume} {349}},\ \bibinfo {pages} {613}
  (\bibinfo {year} {2015})}\BibitemShut {NoStop}%
\bibitem [{\citenamefont {Lv}\ \emph {et~al.}(2015)\citenamefont {Lv},
  \citenamefont {Weng}, \citenamefont {Fu}, \citenamefont {Wang}, \citenamefont
  {Miao}, \citenamefont {Ma}, \citenamefont {Richard}, \citenamefont {Huang},
  \citenamefont {Zhao}, \citenamefont {Chen}, \citenamefont {Fang},
  \citenamefont {Dai}, \citenamefont {Qian},\ and\ \citenamefont
  {Ding}}]{CASTaAs}%
  \BibitemOpen
  \bibfield  {author} {\bibinfo {author} {\bibfnamefont {B.~Q.}\ \bibnamefont
  {Lv}}, \bibinfo {author} {\bibfnamefont {H.~M.}\ \bibnamefont {Weng}},
  \bibinfo {author} {\bibfnamefont {B.~B.}\ \bibnamefont {Fu}}, \bibinfo
  {author} {\bibfnamefont {X.~P.}\ \bibnamefont {Wang}}, \bibinfo {author}
  {\bibfnamefont {H.}~\bibnamefont {Miao}}, \bibinfo {author} {\bibfnamefont
  {J.}~\bibnamefont {Ma}}, \bibinfo {author} {\bibfnamefont {P.}~\bibnamefont
  {Richard}}, \bibinfo {author} {\bibfnamefont {X.~C.}\ \bibnamefont {Huang}},
  \bibinfo {author} {\bibfnamefont {L.~X.}\ \bibnamefont {Zhao}}, \bibinfo
  {author} {\bibfnamefont {G.~F.}\ \bibnamefont {Chen}}, \bibinfo {author}
  {\bibfnamefont {Z.}~\bibnamefont {Fang}}, \bibinfo {author} {\bibfnamefont
  {X.}~\bibnamefont {Dai}}, \bibinfo {author} {\bibfnamefont {T.}~\bibnamefont
  {Qian}}, \ and\ \bibinfo {author} {\bibfnamefont {H.}~\bibnamefont {Ding}},\
  }\href {\doibase 10.1103/PhysRevX.5.031013} {\bibfield  {journal} {\bibinfo
  {journal} {Phys. Rev. X}\ }\textbf {\bibinfo {volume} {5}},\ \bibinfo {pages}
  {031013} (\bibinfo {year} {2015})}\BibitemShut {NoStop}%
\bibitem [{\citenamefont {Yang}\ \emph {et~al.}(2015)\citenamefont {Yang},
  \citenamefont {Liu}, \citenamefont {Sun}, \citenamefont {Peng}, \citenamefont
  {Yang}, \citenamefont {Zhang}, \citenamefont {Zhou}, \citenamefont {Zhang},
  \citenamefont {Guo}, \citenamefont {Rahn}, \citenamefont {Prabhakaran},
  \citenamefont {Hussain}, \citenamefont {Mo}, \citenamefont {Felser},
  \citenamefont {Yan},\ and\ \citenamefont {Chen}}]{YangTaAs}%
  \BibitemOpen
  \bibfield  {author} {\bibinfo {author} {\bibfnamefont {L.~X.}\ \bibnamefont
  {Yang}}, \bibinfo {author} {\bibfnamefont {Z.~K.}\ \bibnamefont {Liu}},
  \bibinfo {author} {\bibfnamefont {Y.}~\bibnamefont {Sun}}, \bibinfo {author}
  {\bibfnamefont {H.}~\bibnamefont {Peng}}, \bibinfo {author} {\bibfnamefont
  {H.~F.}\ \bibnamefont {Yang}}, \bibinfo {author} {\bibfnamefont
  {T.}~\bibnamefont {Zhang}}, \bibinfo {author} {\bibfnamefont
  {B.}~\bibnamefont {Zhou}}, \bibinfo {author} {\bibfnamefont {Y.}~\bibnamefont
  {Zhang}}, \bibinfo {author} {\bibfnamefont {Y.~F.}\ \bibnamefont {Guo}},
  \bibinfo {author} {\bibfnamefont {M.}~\bibnamefont {Rahn}}, \bibinfo {author}
  {\bibfnamefont {D.}~\bibnamefont {Prabhakaran}}, \bibinfo {author}
  {\bibfnamefont {Z.}~\bibnamefont {Hussain}}, \bibinfo {author} {\bibfnamefont
  {S.-K.}\ \bibnamefont {Mo}}, \bibinfo {author} {\bibfnamefont
  {C.}~\bibnamefont {Felser}}, \bibinfo {author} {\bibfnamefont
  {B.}~\bibnamefont {Yan}}, \ and\ \bibinfo {author} {\bibfnamefont {Y.~L.}\
  \bibnamefont {Chen}},\ }\href {\doibase 10.1038/nphys3425} {\bibfield
  {journal} {\bibinfo  {journal} {Nat. Phys.}\ }\textbf {\bibinfo {volume}
  {11}},\ \bibinfo {pages} {728} (\bibinfo {year} {2015})}\BibitemShut
  {NoStop}%
\bibitem [{\citenamefont {Kim}\ \emph {et~al.}(2013)\citenamefont {Kim},
  \citenamefont {Kim}, \citenamefont {Wang}, \citenamefont {Sasaki},
  \citenamefont {Satoh}, \citenamefont {Ohnishi}, \citenamefont {Kitaura},
  \citenamefont {Yang},\ and\ \citenamefont
  {Li}}]{BiSbKimMagnetoTransportExpt}%
  \BibitemOpen
  \bibfield  {author} {\bibinfo {author} {\bibfnamefont {H.-J.}\ \bibnamefont
  {Kim}}, \bibinfo {author} {\bibfnamefont {K.-S.}\ \bibnamefont {Kim}},
  \bibinfo {author} {\bibfnamefont {J.-F.}\ \bibnamefont {Wang}}, \bibinfo
  {author} {\bibfnamefont {M.}~\bibnamefont {Sasaki}}, \bibinfo {author}
  {\bibfnamefont {N.}~\bibnamefont {Satoh}}, \bibinfo {author} {\bibfnamefont
  {A.}~\bibnamefont {Ohnishi}}, \bibinfo {author} {\bibfnamefont
  {M.}~\bibnamefont {Kitaura}}, \bibinfo {author} {\bibfnamefont
  {M.}~\bibnamefont {Yang}}, \ and\ \bibinfo {author} {\bibfnamefont
  {L.}~\bibnamefont {Li}},\ }\href {\doibase 10.1103/PhysRevLett.111.246603}
  {\bibfield  {journal} {\bibinfo  {journal} {Phys. Rev. Lett.}\ }\textbf
  {\bibinfo {volume} {111}},\ \bibinfo {pages} {246603} (\bibinfo {year}
  {2013})}\BibitemShut {NoStop}%
\bibitem [{\citenamefont {Li}\ \emph {et~al.}(2015)\citenamefont {Li},
  \citenamefont {He}, \citenamefont {Lu}, \citenamefont {Zhang}, \citenamefont
  {Liu}, \citenamefont {Ma}, \citenamefont {Fan}, \citenamefont {Shen},\ and\
  \citenamefont {Wang}}]{LiCd3As2NegativeMR}%
  \BibitemOpen
  \bibfield  {author} {\bibinfo {author} {\bibfnamefont {H.}~\bibnamefont
  {Li}}, \bibinfo {author} {\bibfnamefont {H.}~\bibnamefont {He}}, \bibinfo
  {author} {\bibfnamefont {H.-Z.}\ \bibnamefont {Lu}}, \bibinfo {author}
  {\bibfnamefont {H.}~\bibnamefont {Zhang}}, \bibinfo {author} {\bibfnamefont
  {H.}~\bibnamefont {Liu}}, \bibinfo {author} {\bibfnamefont {R.}~\bibnamefont
  {Ma}}, \bibinfo {author} {\bibfnamefont {Z.}~\bibnamefont {Fan}}, \bibinfo
  {author} {\bibfnamefont {S.-Q.}\ \bibnamefont {Shen}}, \ and\ \bibinfo
  {author} {\bibfnamefont {J.}~\bibnamefont {Wang}},\ }\href@noop {} {\bibfield
   {journal} {\bibinfo  {journal} {ArXiv e-prints}\ } (\bibinfo {year}
  {2015})},\ \Eprint {http://arxiv.org/abs/1507.06470} {arXiv:1507.06470}
  \BibitemShut {NoStop}%
\bibitem [{\citenamefont {Liang}\ \emph {et~al.}(2015)\citenamefont {Liang},
  \citenamefont {Gibson}, \citenamefont {Ali}, \citenamefont {Liu},
  \citenamefont {Cava},\ and\ \citenamefont {Ong}}]{LiangCd3As2}%
  \BibitemOpen
  \bibfield  {author} {\bibinfo {author} {\bibfnamefont {T.}~\bibnamefont
  {Liang}}, \bibinfo {author} {\bibfnamefont {Q.}~\bibnamefont {Gibson}},
  \bibinfo {author} {\bibfnamefont {M.~N.}\ \bibnamefont {Ali}}, \bibinfo
  {author} {\bibfnamefont {M.}~\bibnamefont {Liu}}, \bibinfo {author}
  {\bibfnamefont {R.}~\bibnamefont {Cava}}, \ and\ \bibinfo {author}
  {\bibfnamefont {N.}~\bibnamefont {Ong}},\ }\href {\doibase 10.1038/nmat4143}
  {\bibfield  {journal} {\bibinfo  {journal} {Nat. Mater.}\ }\textbf {\bibinfo
  {volume} {14}},\ \bibinfo {pages} {280} (\bibinfo {year} {2015})}\BibitemShut
  {NoStop}%
\bibitem [{\citenamefont {He}\ \emph {et~al.}(2014)\citenamefont {He},
  \citenamefont {Hong}, \citenamefont {Dong}, \citenamefont {Pan},
  \citenamefont {Zhang}, \citenamefont {Zhang},\ and\ \citenamefont
  {Li}}]{HeCd3As2}%
  \BibitemOpen
  \bibfield  {author} {\bibinfo {author} {\bibfnamefont {L.~P.}\ \bibnamefont
  {He}}, \bibinfo {author} {\bibfnamefont {X.~C.}\ \bibnamefont {Hong}},
  \bibinfo {author} {\bibfnamefont {J.~K.}\ \bibnamefont {Dong}}, \bibinfo
  {author} {\bibfnamefont {J.}~\bibnamefont {Pan}}, \bibinfo {author}
  {\bibfnamefont {Z.}~\bibnamefont {Zhang}}, \bibinfo {author} {\bibfnamefont
  {J.}~\bibnamefont {Zhang}}, \ and\ \bibinfo {author} {\bibfnamefont {S.~Y.}\
  \bibnamefont {Li}},\ }\href {\doibase 10.1103/PhysRevLett.113.246402}
  {\bibfield  {journal} {\bibinfo  {journal} {Phys. Rev. Lett.}\ }\textbf
  {\bibinfo {volume} {113}},\ \bibinfo {pages} {246402} (\bibinfo {year}
  {2014})}\BibitemShut {NoStop}%
\bibitem [{\citenamefont {Xiong}\ \emph {et~al.}(2015)\citenamefont {Xiong},
  \citenamefont {Kushwaha}, \citenamefont {Liang}, \citenamefont {Krizan},
  \citenamefont {Wang}, \citenamefont {Cava},\ and\ \citenamefont
  {Ong}}]{XiongCurrentPlume}%
  \BibitemOpen
  \bibfield  {author} {\bibinfo {author} {\bibfnamefont {J.}~\bibnamefont
  {Xiong}}, \bibinfo {author} {\bibfnamefont {S.~K.}\ \bibnamefont {Kushwaha}},
  \bibinfo {author} {\bibfnamefont {T.}~\bibnamefont {Liang}}, \bibinfo
  {author} {\bibfnamefont {J.~W.}\ \bibnamefont {Krizan}}, \bibinfo {author}
  {\bibfnamefont {W.}~\bibnamefont {Wang}}, \bibinfo {author} {\bibfnamefont
  {R.}~\bibnamefont {Cava}}, \ and\ \bibinfo {author} {\bibfnamefont
  {N.}~\bibnamefont {Ong}},\ }\href@noop {} {\bibfield  {journal} {\bibinfo
  {journal} {ArXiv e-prints}\ } (\bibinfo {year} {2015})},\ \Eprint
  {http://arxiv.org/abs/1503.08179} {arXiv:1503.08179} \BibitemShut {NoStop}%
\bibitem [{\citenamefont {Zhang}\ \emph
  {et~al.}(2015{\natexlab{a}})\citenamefont {Zhang}, \citenamefont {Yuan},
  \citenamefont {Xu}, \citenamefont {Lin}, \citenamefont {Tong}, \citenamefont
  {Hasan}, \citenamefont {Wang}, \citenamefont {Zhang},\ and\ \citenamefont
  {Jia}}]{ZhangTaAsTransport}%
  \BibitemOpen
  \bibfield  {author} {\bibinfo {author} {\bibfnamefont {C.}~\bibnamefont
  {Zhang}}, \bibinfo {author} {\bibfnamefont {Z.}~\bibnamefont {Yuan}},
  \bibinfo {author} {\bibfnamefont {S.}~\bibnamefont {Xu}}, \bibinfo {author}
  {\bibfnamefont {Z.}~\bibnamefont {Lin}}, \bibinfo {author} {\bibfnamefont
  {B.}~\bibnamefont {Tong}}, \bibinfo {author} {\bibfnamefont {M.~Z.}\
  \bibnamefont {Hasan}}, \bibinfo {author} {\bibfnamefont {J.}~\bibnamefont
  {Wang}}, \bibinfo {author} {\bibfnamefont {C.}~\bibnamefont {Zhang}}, \ and\
  \bibinfo {author} {\bibfnamefont {S.}~\bibnamefont {Jia}},\ }\href
  {http://arxiv.org/abs/1502.00251} {\bibfield  {journal} {\bibinfo  {journal}
  {arXiv e-prints}\ } (\bibinfo {year} {2015}{\natexlab{a}})},\ \Eprint
  {http://arxiv.org/abs/1502.00251} {arXiv:1502.00251} \BibitemShut {NoStop}%
\bibitem [{\citenamefont {Moll}\ \emph
  {et~al.}(2015{\natexlab{a}})\citenamefont {Moll}, \citenamefont {Potter},
  \citenamefont {Ramshaw}, \citenamefont {Modic}, \citenamefont {Riggs},
  \citenamefont {Zeng}, \citenamefont {Ghimire}, \citenamefont {Bauer},
  \citenamefont {Kealhofer}, \citenamefont {Nair}, \citenamefont {Ronning},\
  and\ \citenamefont {Analytis}}]{MollNbAs}%
  \BibitemOpen
  \bibfield  {author} {\bibinfo {author} {\bibfnamefont {P.~J.}\ \bibnamefont
  {Moll}}, \bibinfo {author} {\bibfnamefont {A.~C.}\ \bibnamefont {Potter}},
  \bibinfo {author} {\bibfnamefont {B.}~\bibnamefont {Ramshaw}}, \bibinfo
  {author} {\bibfnamefont {K.}~\bibnamefont {Modic}}, \bibinfo {author}
  {\bibfnamefont {S.}~\bibnamefont {Riggs}}, \bibinfo {author} {\bibfnamefont
  {B.}~\bibnamefont {Zeng}}, \bibinfo {author} {\bibfnamefont {N.~J.}\
  \bibnamefont {Ghimire}}, \bibinfo {author} {\bibfnamefont {E.~D.}\
  \bibnamefont {Bauer}}, \bibinfo {author} {\bibfnamefont {R.}~\bibnamefont
  {Kealhofer}}, \bibinfo {author} {\bibfnamefont {N.}~\bibnamefont {Nair}},
  \bibinfo {author} {\bibfnamefont {F.}~\bibnamefont {Ronning}}, \ and\
  \bibinfo {author} {\bibfnamefont {J.~G.}\ \bibnamefont {Analytis}},\ }\href
  {http://arxiv.org/abs/1507.06981} {\bibfield  {journal} {\bibinfo  {journal}
  {arXiv e-prints}\ } (\bibinfo {year} {2015}{\natexlab{a}})},\ \Eprint
  {http://arxiv.org/abs/1507.06981} {arXiv:1507.06981} \BibitemShut {NoStop}%
\bibitem [{\citenamefont {Liu}\ \emph {et~al.}(2015)\citenamefont {Liu},
  \citenamefont {Zhang}, \citenamefont {Yuan}, \citenamefont {Lei},
  \citenamefont {Wang}, \citenamefont {Sante}, \citenamefont {Narayan},
  \citenamefont {He}, \citenamefont {Picozzi}, \citenamefont {Sanvito},
  \citenamefont {Che},\ and\ \citenamefont {Xiu}}]{LiuGatedCd3As2}%
  \BibitemOpen
  \bibfield  {author} {\bibinfo {author} {\bibfnamefont {Y.}~\bibnamefont
  {Liu}}, \bibinfo {author} {\bibfnamefont {C.}~\bibnamefont {Zhang}}, \bibinfo
  {author} {\bibfnamefont {X.}~\bibnamefont {Yuan}}, \bibinfo {author}
  {\bibfnamefont {T.}~\bibnamefont {Lei}}, \bibinfo {author} {\bibfnamefont
  {C.}~\bibnamefont {Wang}}, \bibinfo {author} {\bibfnamefont {D.~D.}\
  \bibnamefont {Sante}}, \bibinfo {author} {\bibfnamefont {A.}~\bibnamefont
  {Narayan}}, \bibinfo {author} {\bibfnamefont {L.}~\bibnamefont {He}},
  \bibinfo {author} {\bibfnamefont {S.}~\bibnamefont {Picozzi}}, \bibinfo
  {author} {\bibfnamefont {S.}~\bibnamefont {Sanvito}}, \bibinfo {author}
  {\bibfnamefont {R.}~\bibnamefont {Che}}, \ and\ \bibinfo {author}
  {\bibfnamefont {F.}~\bibnamefont {Xiu}},\ }\href {\doibase
  10.1038/am.2015.110} {\bibfield  {journal} {\bibinfo  {journal} {NPG Asia
  Materials}\ }\textbf {\bibinfo {volume} {7}},\ \bibinfo {pages} {e221}
  (\bibinfo {year} {2015})}\BibitemShut {NoStop}%
\bibitem [{\citenamefont {Potter}\ \emph {et~al.}(2014)\citenamefont {Potter},
  \citenamefont {Kimchi},\ and\ \citenamefont
  {Vishwanath}}]{PotterFermiArcOsc}%
  \BibitemOpen
  \bibfield  {author} {\bibinfo {author} {\bibfnamefont {A.~C.}\ \bibnamefont
  {Potter}}, \bibinfo {author} {\bibfnamefont {I.}~\bibnamefont {Kimchi}}, \
  and\ \bibinfo {author} {\bibfnamefont {A.}~\bibnamefont {Vishwanath}},\
  }\href {\doibase 10.1038/ncomms6161} {\bibfield  {journal} {\bibinfo
  {journal} {Nat. Commun.}\ }\textbf {\bibinfo {volume} {5}},\ \bibinfo {pages}
  {5161} (\bibinfo {year} {2014})}\BibitemShut {NoStop}%
\bibitem [{\citenamefont {Zhang}\ \emph
  {et~al.}(2015{\natexlab{b}})\citenamefont {Zhang}, \citenamefont {Bulmash},
  \citenamefont {Hosur}, \citenamefont {Potter},\ and\ \citenamefont
  {Vishwanath}}]{ZhangWSMQOs}%
  \BibitemOpen
  \bibfield  {author} {\bibinfo {author} {\bibfnamefont {Y.}~\bibnamefont
  {Zhang}}, \bibinfo {author} {\bibfnamefont {D.}~\bibnamefont {Bulmash}},
  \bibinfo {author} {\bibfnamefont {P.}~\bibnamefont {Hosur}}, \bibinfo
  {author} {\bibfnamefont {A.~C.}\ \bibnamefont {Potter}}, \ and\ \bibinfo
  {author} {\bibfnamefont {A.}~\bibnamefont {Vishwanath}},\ }\href@noop {}
  {\bibfield  {journal} {\bibinfo  {journal} {ArXiv e-prints}\ } (\bibinfo
  {year} {2015}{\natexlab{b}})},\ \Eprint {http://arxiv.org/abs/1512.06133}
  {arXiv:1512.06133} \BibitemShut {NoStop}%
\bibitem [{\citenamefont {Moll}\ \emph
  {et~al.}(2015{\natexlab{b}})\citenamefont {Moll}, \citenamefont {Nair},
  \citenamefont {Helm}, \citenamefont {Potter}, \citenamefont {Kimchi},
  \citenamefont {Vishwanath},\ and\ \citenamefont {Analytis}}]{MollCd3As2Osc}%
  \BibitemOpen
  \bibfield  {author} {\bibinfo {author} {\bibfnamefont {P.~J.}\ \bibnamefont
  {Moll}}, \bibinfo {author} {\bibfnamefont {N.~L.}\ \bibnamefont {Nair}},
  \bibinfo {author} {\bibfnamefont {T.}~\bibnamefont {Helm}}, \bibinfo {author}
  {\bibfnamefont {A.~C.}\ \bibnamefont {Potter}}, \bibinfo {author}
  {\bibfnamefont {I.}~\bibnamefont {Kimchi}}, \bibinfo {author} {\bibfnamefont
  {A.}~\bibnamefont {Vishwanath}}, \ and\ \bibinfo {author} {\bibfnamefont
  {J.~G.}\ \bibnamefont {Analytis}},\ }\href@noop {} {\bibfield  {journal}
  {\bibinfo  {journal} {arXiv preprints}\ } (\bibinfo {year}
  {2015}{\natexlab{b}})},\ \Eprint {http://arxiv.org/abs/1505.02817}
  {arXiv:1505.02817} \BibitemShut {NoStop}%
\bibitem [{\citenamefont {Gorbar}\ \emph {et~al.}(2014)\citenamefont {Gorbar},
  \citenamefont {Miransky}, \citenamefont {Shovkovy},\ and\ \citenamefont
  {Sukhachov}}]{GorbarWSMQOs}%
  \BibitemOpen
  \bibfield  {author} {\bibinfo {author} {\bibfnamefont {E.~V.}\ \bibnamefont
  {Gorbar}}, \bibinfo {author} {\bibfnamefont {V.~A.}\ \bibnamefont
  {Miransky}}, \bibinfo {author} {\bibfnamefont {I.~A.}\ \bibnamefont
  {Shovkovy}}, \ and\ \bibinfo {author} {\bibfnamefont {P.~O.}\ \bibnamefont
  {Sukhachov}},\ }\href {\doibase 10.1103/PhysRevB.90.115131} {\bibfield
  {journal} {\bibinfo  {journal} {Phys. Rev. B}\ }\textbf {\bibinfo {volume}
  {90}},\ \bibinfo {pages} {115131} (\bibinfo {year} {2014})}\BibitemShut
  {NoStop}%
\bibitem [{\citenamefont {Gusynin}\ \emph {et~al.}(1996)\citenamefont
  {Gusynin}, \citenamefont {Miransky},\ and\ \citenamefont
  {Shovkovy}}]{Gusynin1996}%
  \BibitemOpen
  \bibfield  {author} {\bibinfo {author} {\bibfnamefont {V.}~\bibnamefont
  {Gusynin}}, \bibinfo {author} {\bibfnamefont {V.}~\bibnamefont {Miransky}}, \
  and\ \bibinfo {author} {\bibfnamefont {I.}~\bibnamefont {Shovkovy}},\ }\href
  {\doibase 10.1016/0550-3213(96)00021-1} {\bibfield  {journal} {\bibinfo
  {journal} {Nucl. Phys. B}\ }\textbf {\bibinfo {volume} {462}},\ \bibinfo
  {pages} {249} (\bibinfo {year} {1996})}\BibitemShut {NoStop}%
\bibitem [{Note1()}]{Note1}%
  \BibitemOpen
  \bibinfo {note} {Recent work\cite {ZhangWSMQOs} shows that the results in
  Ref. \protect \rev@citealpnum {PotterFermiArcOsc} are valid only when the
  Fermi arcs enclose zero area when the chemical potential is at the Weyl
  nodes, and gives a more general result. The results of the present paper in
  the $\theta \rightarrow \pi /2$ limit are consistent with the
  generalization.}\BibitemShut {Stop}%
\end{thebibliography}%

\end{document}